%% file: ms.tex
\documentclass{emulateapj}% [apj,letter]
\slugcomment{{\sc Accepted to ApJ}}

\shortauthors{Heinis et al.}
\shorttitle{Clustering and Star Formation History}
\usepackage{natbib}
\usepackage{multirow}
\bibpunct{(}{)}{;}{a}{}{,}

\begin{document}
\title{Spatial Clustering from GALEX-SDSS samples: Star Formation
  History and large-scale clustering}

\author
    {
      S\'ebastien Heinis\altaffilmark{1},
      Tam\'as Budav\'ari\altaffilmark{1},
      Alex S. Szalay\altaffilmark{1},
      St\'ephane Arnouts\altaffilmark{2},
      Miguel A. Arag\'on-Calvo\altaffilmark{1},
      Ted K. Wyder\altaffilmark{3},
      Tom A. Barlow\altaffilmark{3},
      Karl Foster\altaffilmark{3},
      Peter G.Friedman\altaffilmark{3},
      D. Christopher Martin\altaffilmark{3},
      Patrick Morrissey\altaffilmark{3}, 
      Susan G. Neff\altaffilmark{4},
      David Schiminovich\altaffilmark{5},
      Mark Seibert\altaffilmark{6},
      Luciana Bianchi\altaffilmark{1}, 
      Jos\'e Donas\altaffilmark{7},
      Timothy M. Heckman\altaffilmark{1},
      Young-Wook Lee\altaffilmark{8}, 
      Barry F. Madore\altaffilmark{6},
      Bruno Milliard\altaffilmark{7},
      R. Michael Rich\altaffilmark{9}, 
      and Sukyoung K. Yi\altaffilmark{8}
    }
    \altaffiltext{1}{Department of Physics and Astronomy, The Johns Hopkins
      University, Homewood Campus, Baltimore, MD 21218, USA} 
    \altaffiltext{2}{Canada-France-Hawaii Telescope Corporation, Kamuela, 
      HI-96743, USA}
    \altaffiltext{3}{California Institute of Technology, MC 405-47, 1200
      East California Boulevard, Pasadena, CA 91125, USA}
    \altaffiltext{4}{Laboratory for Astronomy and Solar Physics, NASA
      Goddard Space Flight Center, Greenbelt, MD 20771, USA}
    \altaffiltext{5}{Department of Astronomy, Columbia University, New
      York, New York 10027, USA}
    \altaffiltext{6}{Observatories of the Carnegie Institution of
      Washington, 813 Santa Barbara St., Pasadena, CA 91101, USA}
    \altaffiltext{7}{Laboratoire d'Astrophysique de Marseille - CNRS,
      P\^ole de l'\'Etoile Site de Ch\^ateau-Gombert, 38, rue
      Fr\'ed\'eric Joliot-Curie 13388 Marseille cedex 13, France}
    \altaffiltext{8}{Center for Space Astrophysics, Yonsei University,
      Seoul 120-749, Korea}
    \altaffiltext{9}{Department of Physics and Astronomy, University
      of California, Los Angeles, CA 90095, USA}

\begin{abstract}
  We measure the projected spatial correlation function $w_p(r_p)$
  from a large sample combining GALEX ultraviolet imaging with the
  SDSS spectroscopic sample. We study the dependence of the clustering
  strength for samples selected on $(NUV - r)_{abs}$ color, specific
  star formation rate (SSFR), and stellar mass.  We find that there is
  a smooth transition in the clustering of galaxies as a function of
  this color from weak clustering among blue galaxies to stronger
  clustering for red galaxies. The clustering of galaxies within the
  ``green valley'' has an intermediate strength, and is consistent
  with that expected from galaxy groups. The results are robust to the
  correction for dust extinction. The comparison with simple
  analytical modeling suggests that the halo occupation number
  increases with older star formation epochs. When splitting according
  to SSFR, we find that the SSFR is a more sensitive tracer of
  environment than stellar mass.
\end{abstract}

\maketitle
\section{Introduction}
The nature of star forming galaxies has deeply changed with the
evolution of the Universe. Besides the strong decrease in the cosmic
star formation rate since $z\sim 1$ \citep[e.g.,][]{Hopkins_2006,
  Schiminovich_2005}, the bulk of star formation has shifted from high
stellar mass to low stellar mass systems, \citep[``downsizing'',
see][]{Bundy_2006, Cowie_1996, Juneau_2005}, a phenomenon which
exhibits a similar trend in dark matter halo mass
\citep{Heinis_2007}. In the local Universe results from optical
surveys show that the galaxy population is bimodal, divided at a
stellar mass of $\sim 3\times 10^{10} M_{\odot}$ between actively
star-forming blue sequence galaxies and passively evolving red
sequence galaxies \citep{Kauffmann_2003a}.  This bimodality in galaxy
properties has been observed from low redshift up to $z\sim 1.5$
\citep{Bell_2004, Cirasulo_2007, Cassata_2008, Franzetti_2007}
although the proportion between the red and blue sequence has been
changing. Several studies indeed show that the density of red galaxies
has been increasing since $z = 2$ while the density of galaxies on the
blue sequences remained roughly constant \citep{Bell_2004,
  Faber_2007}. These are further pieces of evidence that some
previously star-forming galaxies have quenched their star formation
and moved from the blue to the red sequence at a rate balanced by
ongoing star formation among the remaining blue sequence galaxies
\citep{Arnouts_2007, Martin_2007}. Various mechanisms can be involved
in the quenching of star formation. Galaxy mergers can trigger AGN
activity \citep[e.g.,][]{Hopkins_2007}, and hence AGN feedback, which
is efficient in further preventing gas cooling \citep{Croton_2006}.
Supernov\ae~explosions inject energy into the interstellar medium and
may expel gas from galaxies in low-mass halos
\citep[e.g.,][]{Dekel_1986}. Galaxy environment is also an important
factor in this evolution: in clusters for instance, several processes
such as starvation \citep{Larson_1980}, harassment \citep{Moore_1999}
or ram pressure stripping \citep{Gunn_1972}, which can be effective at
large distances from the cluster center \citep{Chung_2007,
  Tonnesen_2007}, may combine to lead to morphological transformations
and removal of the gas reservoir. Numerical studies also suggest that
accounting for gravitational heating through satellite accretion can
reproduce the observed trends between star formation and environment
\citep{Khochfar_2008}.

A common tool used to place constraints on the physical mechanisms
involved in this evolution has been the study of galaxies' properties
and their link with environment within the color-magnitude
diagram. Recent studies have showed that at fixed luminosity the
fraction of red galaxies increases in denser environments, while the
mean color of the red and blue sequences are fairly insensitive to
local density \citep{Balogh_2004}. \citet{Hogg_2003} found that for
blue galaxies, at fixed star formation history, stellar mass is weakly
correlated with environment, and that luminous and faint red galaxies
reside in denser environments. \citet{Cucciati_2006} showed that the
relation between color and local density evolved between $z=1.5$ and
$z=0$, which suggests that the links between galaxy properties and
environment observed at the present epoch are not only determined by
galaxy formation conditions (``nature'') but also by contributions
from galaxy evolution processes (``nurture'').

Extending these studies using GALEX \citep{Martin_2005} data,
\citet{Wyder_2007} showed that the galaxy distribution is less bimodal
in ultraviolet (UV) optical color-magnitude diagrams, owing to the
greater sensitivity of the UV to star formation and dust. Previous
work also showed that there is a very tight correlation between the
$(NUV - r)_{\rm{abs}}$\footnote{We denote absolute magnitudes by the
  'abs' subscript.} color and star formation history
\citep{Salim_2005}. UV-optical colors furthermore highlight a galaxy
population within the intermediate region between the red and the blue
sequences, the so-called ``green valley'' \citep{Martin_2007}.

\citet{Schawinski_2007} studied early-type galaxies within the
UV-optical color magnitude diagram; they showed that this population
exhibits residual star formation, and that the fraction of UV-bright
early-type galaxies is higher in lower density environments. In this
paper, we extend this study by using the combination of GALEX imaging
with SDSS spectroscopic data to measure the dependence of spatial
clustering on the $(NUV - r)_{\rm{abs}}$ color, in order to link star
formation history with environment, the latter being traced by the
large-scale clustering. We use here a simple power-law approximation
to describe the clustering of the galaxies, in order to estimate the
mass of the host dark matter halos.

The outline of this paper is as follows: in Section \ref{sec_data} we
present the samples from the GALEX surveys, as well as the internal
dust correction procedure. Section \ref{sec_clustering} deals with the
clustering analysis tools, and contains a discussion of the systematic
errors previously noticed for clustering measurements from GALEX data
\citep{Milliard_2007}. We present the results in Section
\ref{sec_results}, discuss them in Section \ref{sec_discussion}, and
conclude in Section \ref{sec_conclusion}. Throughout this paper we use
the concordance $\Lambda$CDM cosmology with $\Omega_m = 0.3$ and
$\Omega_{\Lambda} = 0.7$; we note $H_0 = 100 h$km s$^{-1}$
Mpc$^{-1}$. We use $h=1$ for clustering computations to facilitate
comparisons with other studies, and $h=0.7$ for all absolute
magnitudes derivations.

\section{Data Samples}\label{sec_data}
We consider observations from two different GALEX surveys: the Medium
Imaging Survey (MIS) to use the UV photometry for the overall galaxy
population detected by the SDSS spectroscopic survey, and the All sky
Imaging Survey (AIS), to build a UV selected
sample. \citet{Budavari_2009} described in detail how we built the
catalogs and footprints; we only summarize here the main characteristics.

We computed the unique area observed in the GALEX MIS and AIS. The
fields target pre-defined positions on the sky. These 47,162 points
define disjoint cells by including in each cell those regions of sky
closer to the center of that cell than to the center of any other
cell, similar to the way in which a Voronoi tessellation is
defined. The intersection of the circular field-of-view of a given
field and the targeted cell is the \textit{primary} region for this
field. We use here only objects within the primary resolution of AIS
or MIS fields. The union of the primary region from all fields defines
the primary footprint of a survey. We then use the formalism of
\citet{Budavari_2008} to compute the footprint of the GALEX-SDSS
overlap.

The GALEX 3$^{\rm{rd}}$ Release \citep[GR3;][]{Morrissey_2007} dataset
was cross-matched to the SDSS DR6 \citep{Adelman_2008} using a
matching radius of 4\arcsec.  We used the GALEX sources with one and
only one SDSS counterpart within the search radius. In the following
we only consider SDSS primary sources. We further restricted the SDSS
sources to spectroscopically confirmed galaxies with $z<0.3$ and
$14.5<r_{petro}<17.6$. The bright cut avoids the shredded galaxies,
while the faint selection ensures a homogeneous spectroscopic
completeness over the sky. For GALEX sources, we retained the sources
within 0.5 degrees of the field centers, as the photometry accuracy
decreases beyond this limit, and artifacts concentrate at the edges of
the GALEX field of view \citep{Morrissey_2005}. We censor sources
using GALEX (flagmap values 2 and 4) and SDSS (\texttt{BLEEDING},
\texttt{BRIGHT\_STAR}, \texttt{TRAIL} and \texttt{HOLE}) masks. We use
the intersection of the GALEX and SDSS footprints as described in
\citet{Budavari_2008}. We corrected GALEX apparent magnitudes for
Galactic extinction using the \citet{Schlegel_1998} dust maps and the
formulas described in \citet{Wyder_2007}. We estimated
$k-$corrections\footnote{using LePhare:
  \url{http://www.oamp.fr/people/arnouts/LE\_PHARE.html}} at $z=0$
after fitting the 7-band photometry (FUV, NUV, $u, g, r, i, z$) to a
set of \citet{Bruzual_2003}
templates\footnote{\url{http://www.oamp.fr/people/arnouts/LE\_PHARE/lephare\_sed\_42GISSEL.tar.gz}};
the absolute magnitudes are $k-$corrected to $z = 0$ unless otherwise
stated. We also derived $k-$corrections at $z=0.1$ by shifting the
filters blueward in wavelength by a factor $1/(1+z) = 1/1.1$, in order
to compute dust extinction correction (see \S \ref{sec_dust_cor}).

\subsection{MIS/SDSS selection}\label{sec_data_mis}
\begin{figure}[htbp]
  \plotone{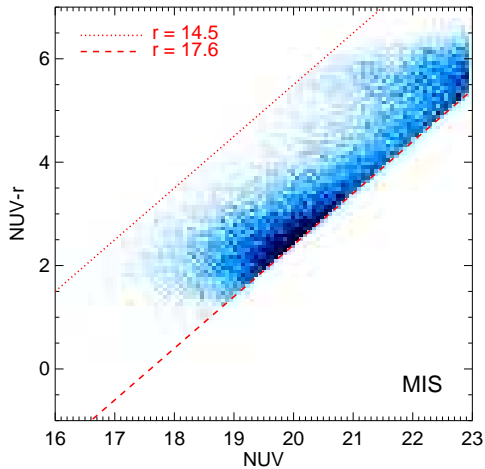}
  \caption{\small Color magnitude diagram for MIS selection. The
  dotted and dashed lines show the cuts in $r$ adopted for SDSS
  spectroscopic objects.}
  \label{fig_mis_sel}
\end{figure}

The MIS observations are deep enough to probe the overall population
of the SDSS Main Galaxy spectroscopic sample \citep{Strauss_2002}. We
used MIS fields with exposure times greater than 1000s for the NUV
band, and objects with NUV $<$23. Fig.\ \ref{fig_mis_sel} shows the
color-magnitude diagram in the (NUV, NUV-r) plane for the MIS
sources. The dashed line corresponds to the faint cut in $r$, and the
dotted line to the bright cut we adopt. This sample is obviously not
UV-selected, as the density decreases for objects fainter than $r =
17.75$, the SDSS target magnitude limit. There is however a clear hint
of a bimodality in color, with a separation around NUV-r $\sim$ 4.75
between the red and the blue sequence. The final sample contains
21,895 galaxies over 598.1 sqdeg.

\subsection{AIS: UV selection}\label{sec_data_ais}
\begin{figure}[htbp]
  \plotone{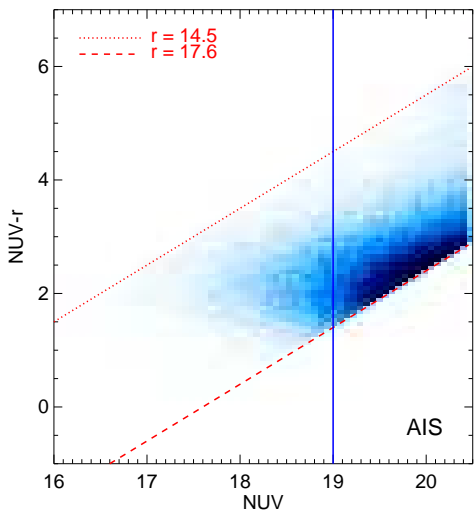}
  \caption{\small Color magnitude diagram for AIS selection. The
    dotted and dashed lines show the cuts in $r$ adopted for SDSS
    spectroscopic objects, and the solid line the NUV cut chosen to
    ensure an UV-selected sample.}
  \label{fig_ais_sel}
\end{figure}

We used AIS observations to build UV-selected samples from the
GALEX-SDSS spectroscopic samples. Figure \ref{fig_ais_sel} shows as
before the color-magnitude diagram for AIS objects, with an additional
dotted line representing the apparent magnitude cut at NUV$=$19 we use
to ensure a UV-selected sample. The fraction of objects brighter than
NUV $= 19$ and fainter than $r = 17.6$ that are lost is 3\%. We note
that the NUV cut removes most of the galaxies on the red sequence; hence
our UV-selected galaxy sample mainly contains blue galaxies. The size
of the sample is 14,351 galaxies, over an area of 3645.8 sqdeg.

\subsection{Dust extinction correction}\label{sec_dust_cor}

In the following we study the dependence of clustering as a function
of the $(NUV - r)_{\rm{abs}}$ color, which correlates tightly with
star formation history and provides a very good separation between
galaxy populations. However interstellar dust absorbs and scatters
stellar light, reddening galaxy colors and interfering with the
interpretation of the colors in terms of the star formation history.
Several recipes are available to correct for dust reddening, but one
often has to rely on various assumptions about the dust content of
galaxies. For instance, galaxies with intermediate colors may have
redder colors than blue galaxies because of dust, or because of older
stellar populations. Hence applying a given dust extinction correction
may imply potential errors for transition or red galaxies. We will
therefore present clustering measurements as a function of both
non-corrected and corrected colors. We use the relation presented by
\citet{Johnson_2006}:

\begin{eqnarray}\label{eq_dust_cor}
  A_{FUV} & = & 1.27 - 1.56x + 1.35y- 1.24xy \\
  x       & = & D_n(4000) -1.25 \nonumber \\
  y       & = & (NUV - r)_{abs,0.1} - 2\nonumber
\end{eqnarray}

\noindent to estimate the extinction correction in the FUV band,
$A_{FUV}$. \citet{Johnson_2006} used the narrowband color $D_n(4000)$
from \citet{Kauffmann_2003a} to calibrate their relation, while we
consider here the $D(4000)$ index directly available from the SDSS DR6
pipeline, which involves the ratio of the flux over a longer wavelength
range. We checked that for objects in common between the SDSS DR6 and the sample used by
\citet{Kauffmann_2003a} that the difference between their estimate of
$D_n(4000)$ and our estimate of $D(4000)$ is reasonably small with a dispersion of $\sim
0.06$. Finally, we did not apply any dust correction to objects without
star formation, defined here as those galaxies with $EW(H\alpha)<0$.
These objects represent 7\% of the MIS/SDSS sample and less than 1\%
of the AIS UV-selected one. Negative values of $A_{FUV}$ according to
eq. \ref{eq_dust_cor} are set to 0.

We used the \citet{Calzetti_2000} extinction law to convert the
extinction in the FUV band into extinction in the NUV, $g$ and $r$
bands with: $A_{NUV} = 0.81A_{FUV}$, $A_{g} = 0.46A_{FUV}$ and $A_{r}
= 0.35A_{FUV}$.

\section{Clustering analysis}\label{sec_clustering}
We chose to quantify galaxy clustering in our samples using the
real-space correlation function. As redshift surveys suffer from
redshift distortions due to peculiar velocities, we followed the
common practice \citep{Fisher_1994} of measuring first the
redshift-space correlation function $\xi(r_p, \pi)$, which is useful
for distinguishing between the contribution of galaxy pairs along the
line of sight $\pi$ and the direction perpendicular to the line of
sight $r_p$. We used the \cite{Landy_1993} estimator:

\begin{equation}
  \xi(r_p, \pi) = \frac{DD - 2DR + RR}{RR}
\end{equation}

\noindent where DD, DR and RR are the data-data,
data-random and random-random pair numbers, respectively. 
To generate random catalogs, we distributed objects randomly
on the sky within the GALEX-SDSS footprint in non-masked areas, which
takes care of the angular selection function. We then randomly picked
galaxy properties (such as redshift and magnitude) that we assigned to
the random objects. This procedure has been shown to provide accurate
results, and enabled us to study the dependence of clustering on galaxy
properties by fully taking into account selection effects
\citep{Li_2006a}. While the GALEX-SDSS overlap is fairly patchy, we
checked that the integral constraint is negligible for this study.

We used 50 times more random objects than the
number of galaxies in the sample under consideration to compute the
correlation function. We found out that this is the minimum factor to
estimate accurately the correlation function on small scales, given
the fairly low statistics of the samples we use here.

As redshift distortions occur only on the line-of-sight, one can get
rid of their effects by computing the projected correlation function

\begin{equation}
  w_p(r_p) = 2\int_{0}^{\pi_{max}} \xi(r_p, \pi)d\pi.
\end{equation}

After checking for convergence of the integration step, we used here
$\pi_{max} = 25 h^{-1}$Mpc. The projected correlation function is
finally related to the real-space correlation function $\xi(r)$ by the
relation \citep{Davis_1983}
\begin{equation}
   w_p(r_p) = 2\int_{0}^{\infty} \xi\left(\sqrt{r_p^2 + y^2}\right)dy
\end{equation}
where y is the real-space distance along the line of sight.

Under the assumption of a power-law $\xi(r) = (r/r_0)^{-\gamma}$ this
relation yields
\begin{equation}
   w_p(r_p) = r_p\left(\frac{r_p}{r_0}\right)^{-\gamma}\frac{\Gamma\left(\frac{1}{2}\right)\Gamma\left(\frac{\gamma-1}{2}\right)}{\Gamma\left(\frac{\gamma}{2}\right)}.
\end{equation}
We will use the parameters $r_0$, the correlation length, and
$\gamma$, the slope of the real-space correlation function, to
describe the clustering of the galaxies. We use jackknife sampling to
derive the error bars on $w_p(r_p)$; this procedure has been shown to
provide accurate error measurements on $w_p(r_p)$
\citep{Zehavi_2005}. Specifically we divided our samples into $N$
separate regions (32 for AIS sample, 20 for MIS), large enough so that
their volume is significant. We then computed $w_p(r_p)$ for the full
sample but removing one jackknife region in turn. For a given bin
$r_{p_i}$, the errors from these $N$ samples are then given by:
\begin{equation}
\sigma^2(w(r_{p_i})) = \frac{N-1}{N}\sum_{j=1}^{N} (w^j(r_{p_i}) - \langle w(r_{p_i})\rangle)^2.
\end{equation}

\subsection{Fiber collision correction}
Due to the finite size of the fibers used to take the SDSS spectra,
fibers can not be set closer than 55\arcsec, resulting in a lack of
close pairs at comoving scales smaller than $\sim 80h^{-1}$ kpc (at $z
= 0.1$). We use a statistical approach to correct from this, following
\citet{Hawkins_2003} and \citet{Li_2006a}.

\begin{figure*}[!t]%
  %\epsscale{2.45}%
  \plottwo{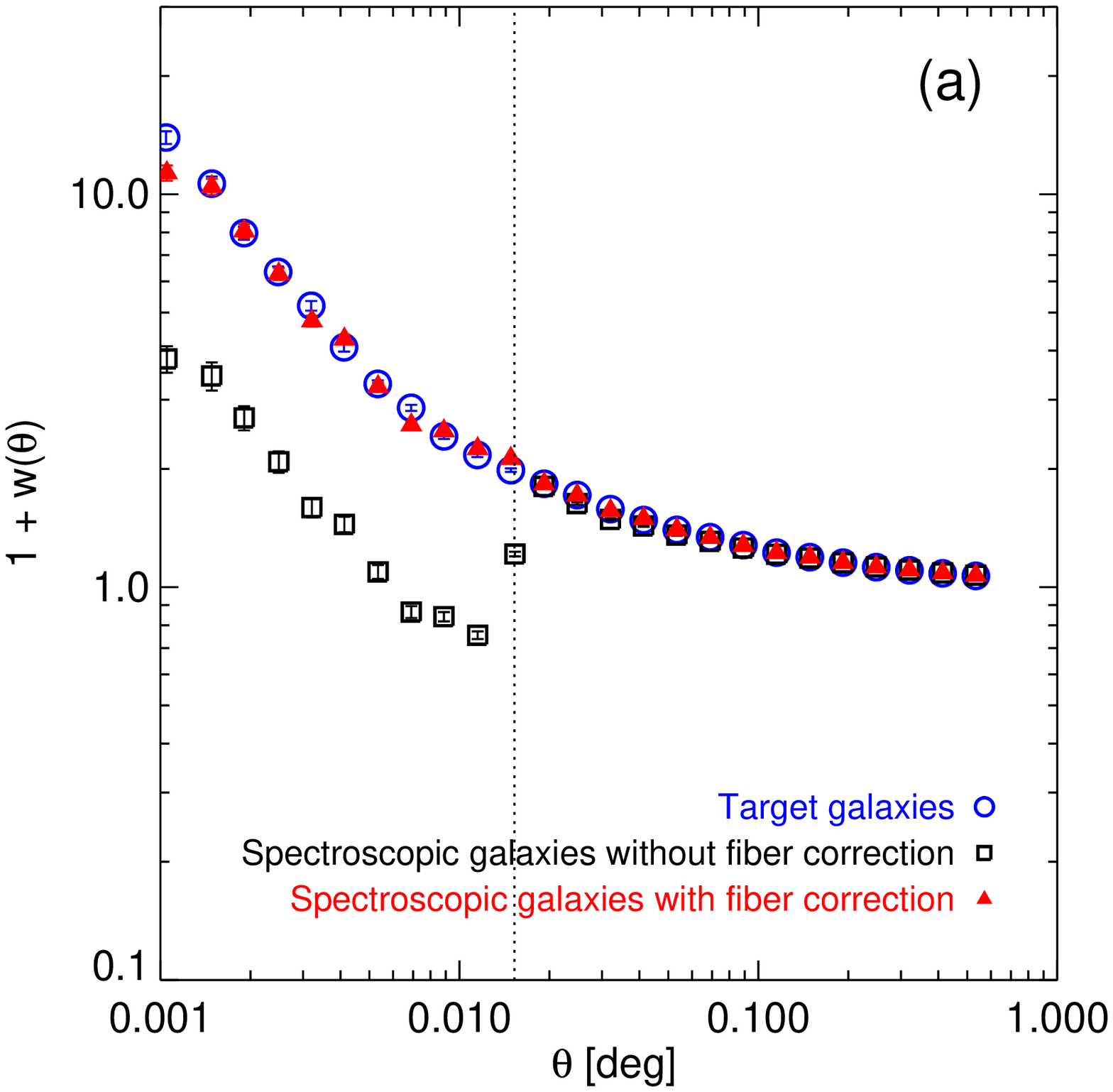}{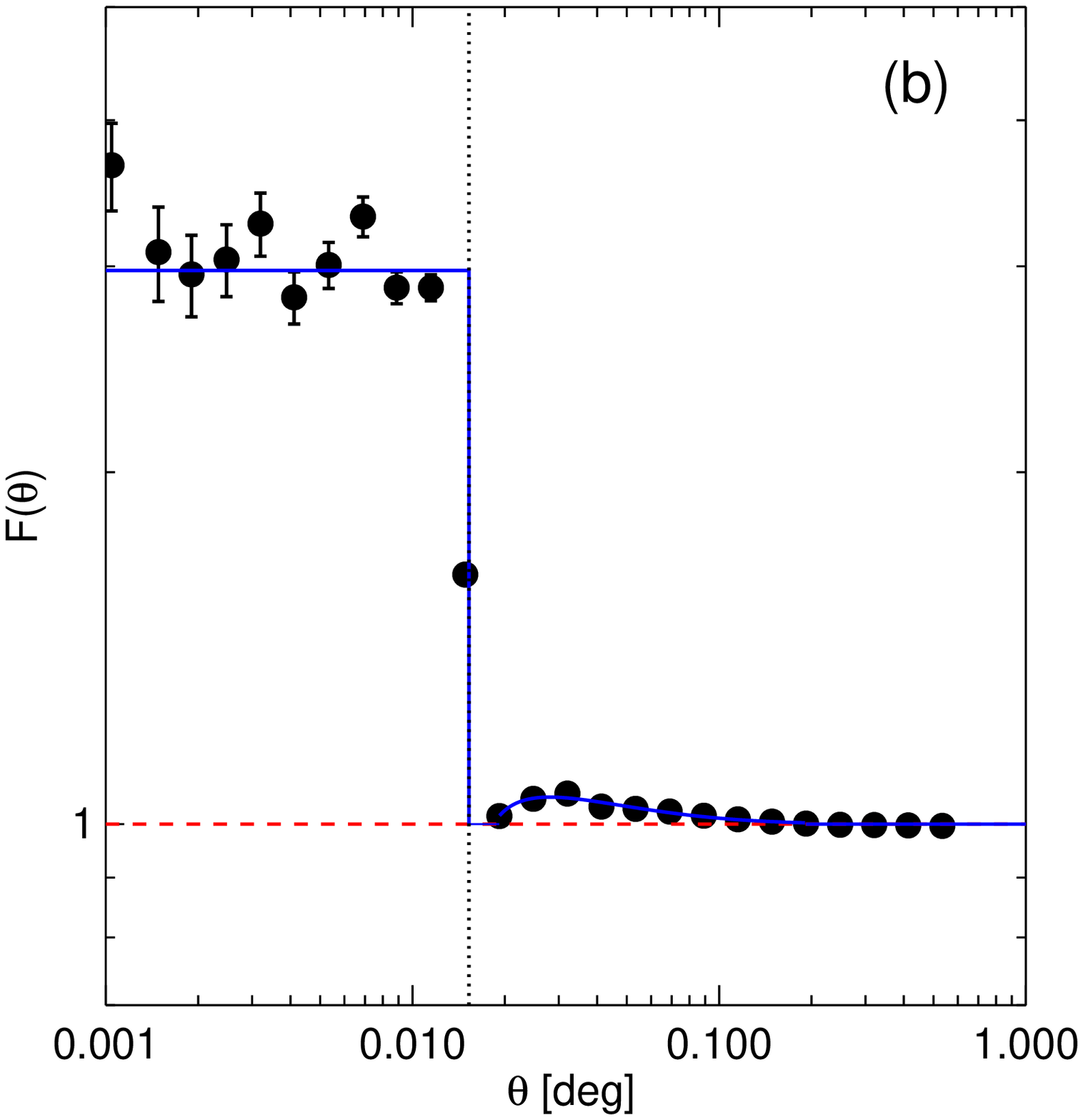}
  \caption{\small Fiber collision correction.\textit{Left:} angular
  correlation function of SDSS target galaxies ($w_t(\theta)$, open
  circles), SDSS sources with assigned galaxy spectra ($w_z(\theta)$,
  open squares). The filled triangles show the angular correlation
  function of the SDSS sources with assigned galaxy spectra computed
  using the correction from fiber collision.\textit{Right:} the
  symbols show the ratio $F(\theta) = [1+w_t(\theta)]/[1+w_z(\theta)]$
  derived from the measurements shown on the left plot. The solid line
  shows the functional form we used for this ratio in order to correct
  for fiber collisions. The dotted line shows the 55\arcsec limit.}
  \label{fig_fiber_col}
\end{figure*}

We measured the angular correlation function $w_t(\theta)$ for the
objects classified as galaxies and selected for spectroscopic
measurement by the SDSS (i.e. target galaxies, open circles on fig.\
\ref{fig_fiber_col}a) and for SDSS objects with assigned galaxy spectra
$w_z(\theta)$ (open squares on fig.\ \ref{fig_fiber_col}a) within the
AIS footprint (large enough to provide an accurate measurement). We
then used the ratio
\begin{equation}
  F(\theta) = \frac{1+w_t(\theta)}{1+w_z(\theta)}
\end{equation}
as an estimate for the correction factor needed to correct for fiber
collision. The measured values of this ratio are shown in fig.\
\ref{fig_fiber_col}b); we used a functional form for this correction
factor represented by the solid lines on fig.\
\ref{fig_fiber_col}b. $F(\theta)$ can be then used to weight the
galaxy pairs according to their angular distance. As a test we
recomputed the angular correlation function of the SDSS objects with
assigned galaxy spectra using this weight (filled triangles on fig.\
\ref{fig_fiber_col}a); the agreement between this result and the
angular correlation function of the target galaxies is excellent.

\subsection{Systematic errors}\label{sec_systematics}
\begin{figure*}[!t]%
  %\epsscale{2.45}%
  \plottwo{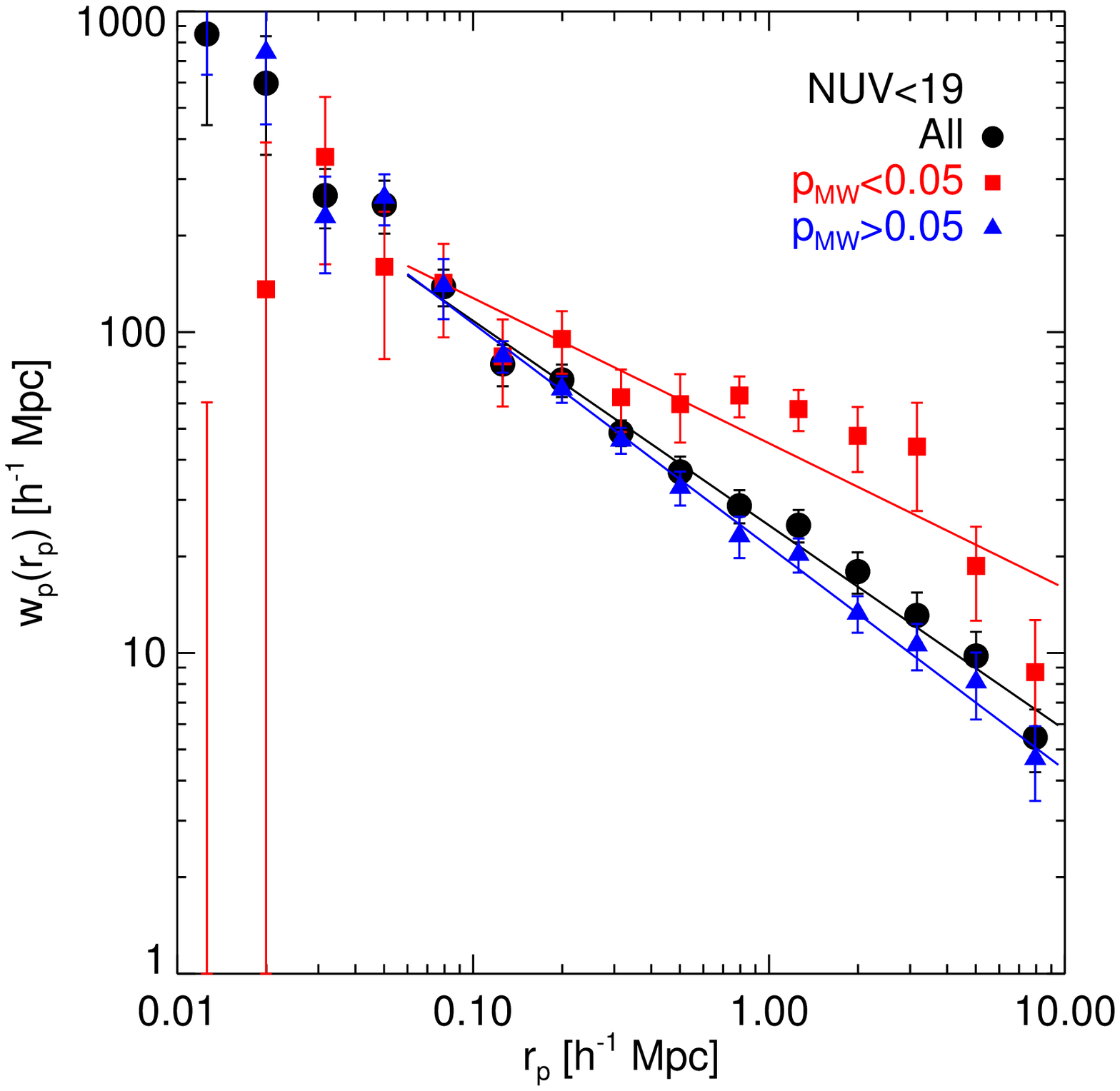}{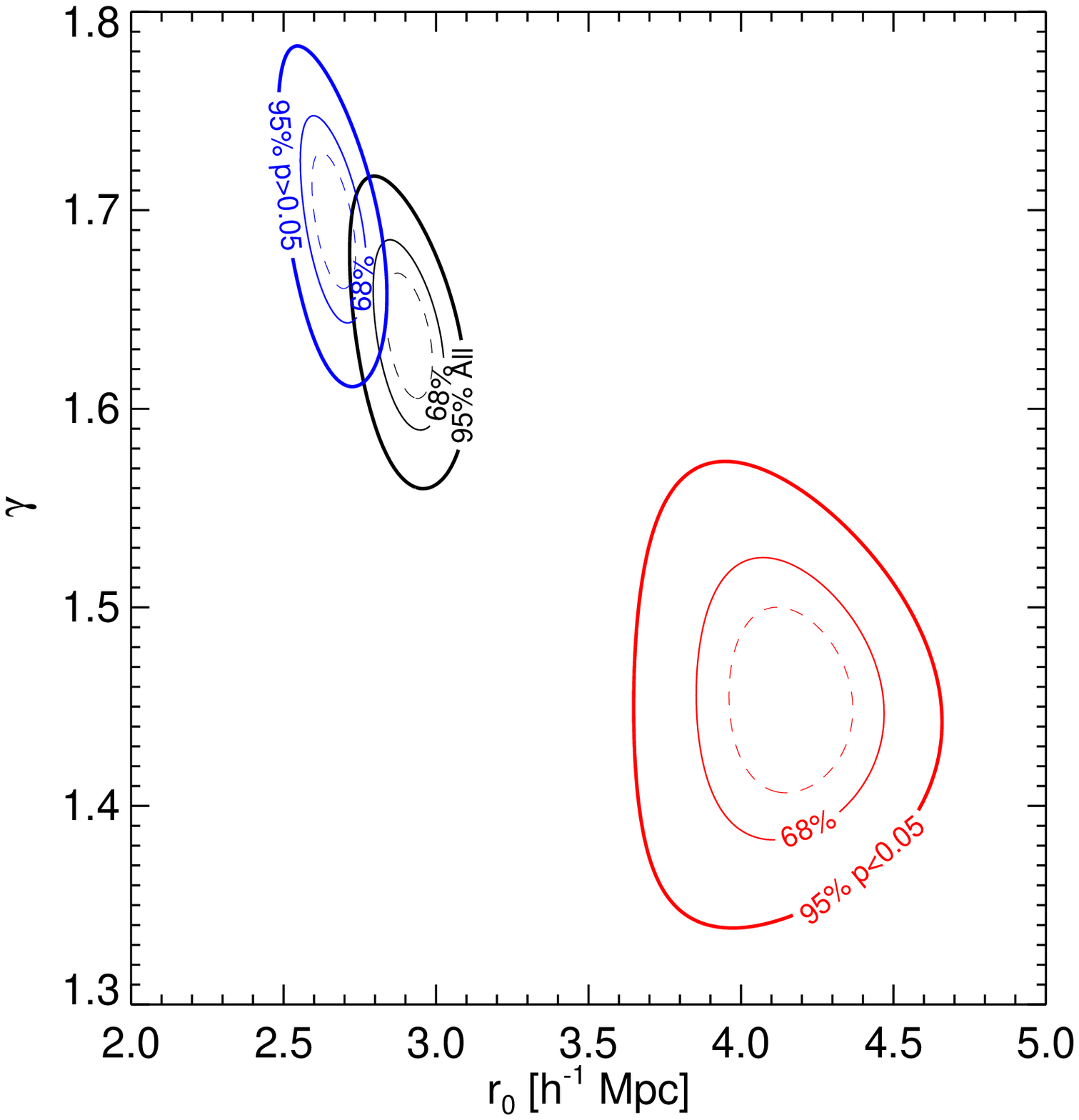}
  \caption{\small \textit{Left:} projected correlation function
    $w_p(r_p)$ from the AIS sample with NUV$<19$. Circles show
    $w_p(r_p)$ for the full sample (All), squares $w_p(r_p)$ for the
    fields flagged as outliers in terms of UV photometry by the
    Mann-Whitney statistics ($p<0.05$), and triangles $w_p(r_p)$ for
    the remaining fields ($p>0.05$). Solid lines represent the best
    power law fits. \textit{Right:} $\chi^2$ contours of the power law
    fits in the ($r_0, \gamma$) plane; the contours labels include the
    legend for each correlation function of the left panel
    plot. Dashed lines show the contour at $\chi^2_{min} + 1$, while
    the thin and thick solid lines show the contours at 68\% and 90\%
    respectively.}
  \label{fig_wrp_mw_ais}
\end{figure*}

In \citet{Milliard_2007} we noticed that the angular correlation
function from GALEX data shows signatures similar to those expected in
the case of low-level field-to-field variations. This prevented us
from estimating the angular correlation function from the whole
dataset at once. The approach we used was to estimate the clustering
from the angular correlation function by measuring $w(\theta)$
independently for each GALEX field, and then averaging the
results. This method cannot be used in the specific case of the
GALEX-SDSS spectroscopic samples, given the low statistics. Hence we
performed a series of tests to see whether the order of magnitude of
the effects on the spatial correlation function are the same as those
on the angular correlation function. As in \citet{Milliard_2007}, we
use the Mann-Whitney statistic to determine which GALEX fields should
be considered as outliers in terms of UV photometry. We consider
potential outliers fields those with a $p$-value from the Mann-Whitney
test lower than 0.05\footnote{This means that the probability that the
  photometry of these fields comes from the same parent distribution
  as the other fields is lower than 0.05.}. We show fig.\
\ref{fig_wrp_mw_ais} (left) the projected correlation functions
$w_p(r_p)$ (computed using all data at once) of the full AIS
sample\footnote{The trends are similar for other apparent magnitude
  cuts and the MIS sample.}  (filled circles), of the fields expected
to be outliers ($p$-value $<0.05$, filled squares, 16\% of the full
sample) and the fields that are expected to have an homogeneous UV
photometry ($p$-value $>0.05$, filled triangles). The correlation
function of the outlier fields is very different from the fields with
a $p$-value $>0.05$, showing a higher amplitude, and a shallower
slope. Fitting these results by power laws (see $\chi^2$ contours on
fig.\ \ref{fig_wrp_mw_ais}, right) shows that the outlier fields have
a correlation length roughly 1.5 times larger than the fields with
high $p$-values. These outlier fields have a fairly small impact on
the clustering parameters derived for the full sample; excluding or
including them from the analysis does not change the conclusions of
this study. Nevertheless, these fields indeed show a very different
clustering. This stronger clustering may be spurious, or may be
real. In the latter case, it might be due to large-scale structure. To
further test this hypothesis, we computed $w_p(r_p)$ from the entire
SDSS data within the AIS footprint, splitting the fields as before,
based on the Mann-Whitney results obtained from UV photometry. The
trends are similar, with the SDSS data within the footprint of the AIS
outlier fields showing a significant stronger
clustering\footnote{Trends are also similar for SDSS data within MIS
  footprint, when splitting fields according to MIS photometry.}. We
also performed the Mann-Whitney test on SDSS data, using the $u$ band;
60\% of the outlier fields are the same as the ones according to NUV
photometry. The clustering derived from the fields classified as
outliers with both $u$ and NUV photometry actually dominate the
clustering signal observed for the outliers on fig.\
\ref{fig_wrp_mw_ais}.

As a final test, a friend-of-friend structure finding algorithm
applied on SDSS spectroscopic data shows that there is a slight trend
for fields with lower $p$-values derived from UV photometry to have
larger galaxy groups, as recovered by the friend-of-friend
algorithm. These tests indicate that the stronger clustering, as
measured by the combination of GALEX and SDSS spectroscopic data,
observed for the fields considered as outliers based on the
Mann-Whitney test applied on UV photometry is genuine. We
checked for instance that the large-scale structure known as the
``Sloan Great Wall'' \citep{Gott_2005} is not within our sample; our
tests rather show that the fields we classified as outliers contain
galaxy groups with larger numbers of members. In consequence, these
fields need to be considered in the global analysis. We hereafter use
the full AIS and MIS samples, and compute the projected correlation
function from the whole dataset.

\section{Results}\label{sec_results}

\subsection{Redshift space clustering}
\begin{figure*}[!t]%
  %\epsscale{2.45}%
  \plottwo{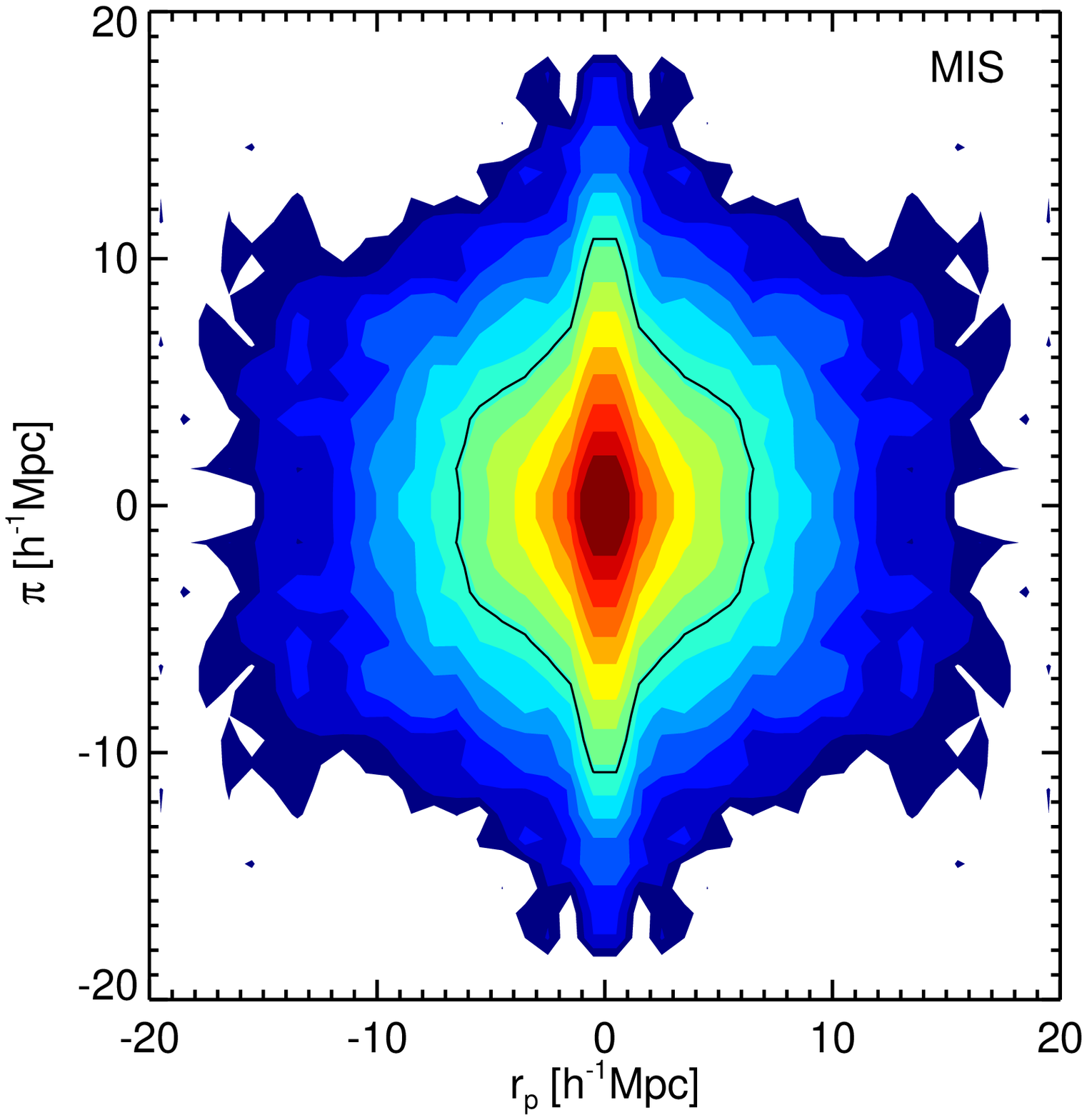}{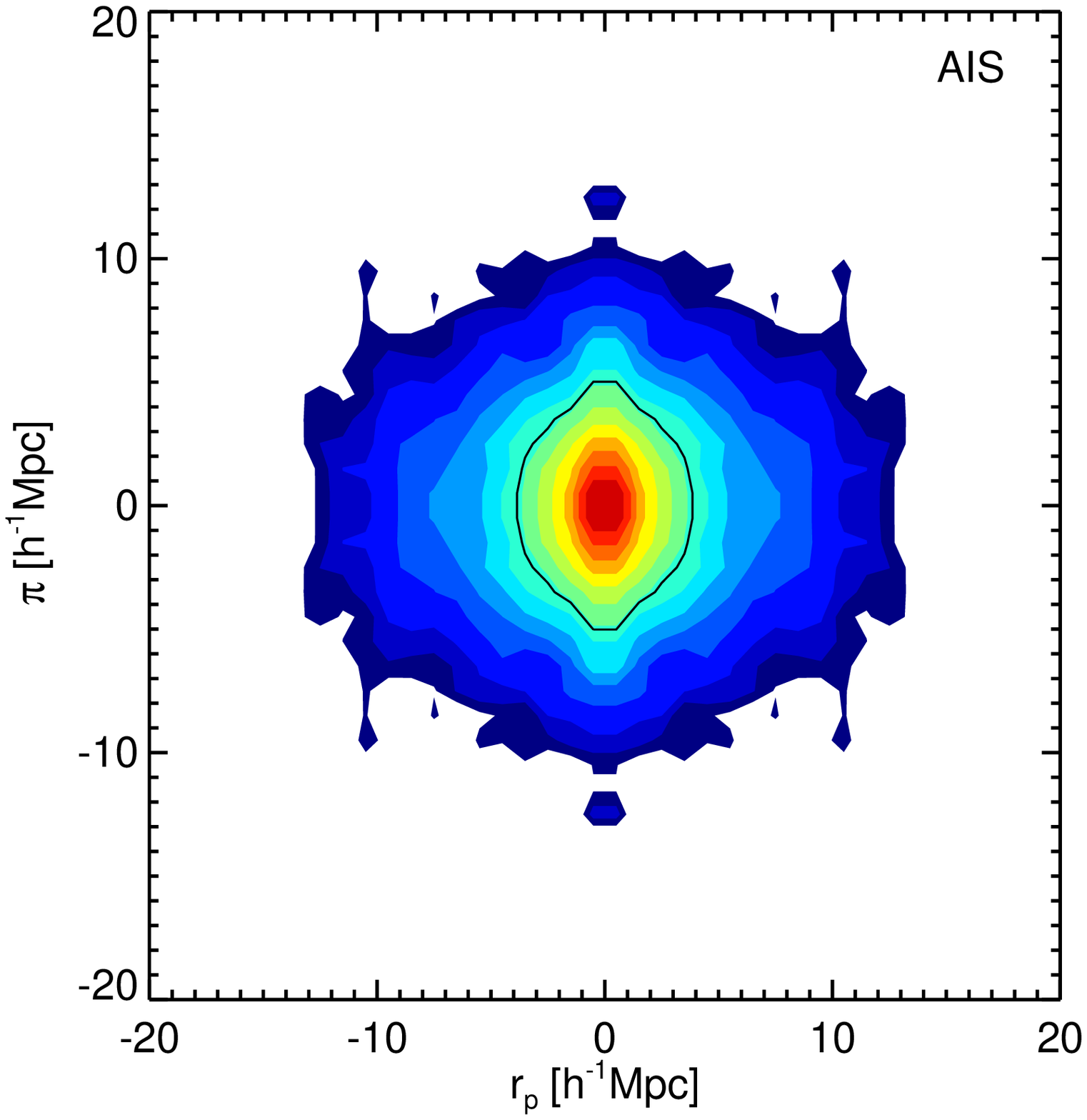}
  \caption{\small Redshift space correlation functions. \textit{Left:}
    Contours of $\xi(r_p, \pi)$ for the MIS sample.\textit{Right:}
    Same as left panel but for the AIS sample. The solid black line
    shows the contour at $\xi(r_p, \pi) = 1$. The contour levels are
    logarithmically spaced.}
  \label{fig_xi_mis_ais}
\end{figure*}

Figure \ref{fig_xi_mis_ais} shows the redshift space correlation
function $\xi(r_p, \pi)$ computed from the MIS sample (left) and the
AIS sample (right). The contours show the well-known effects due to
peculiar velocities \citep{Kaiser_1987}: at small scales, galaxies in
clusters lead to distortions along the line of sight (fingers of God);
this effect is clearly noticed in our MIS/SDSS sample. As expected,
this feature is much less prominent in the UV-selected AIS sample,
which contains very few red galaxies. At large scales, coherent infall
on large structures causes a squashing of the contours, clearly seen
in both samples.

\subsection{Clustering as a function of $(NUV -r)_{\rm{abs}}$
color}\label{sec_res_nuv_r}

\begin{figure}[!t]%[htbp]
  \epsscale{1.}
  \plotone{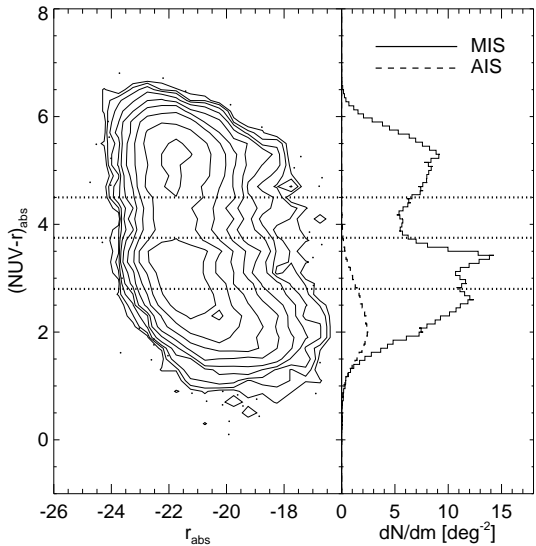}
  \caption{\small \textit{Left:} color-magnitude diagram from the
    MIS/SDSS sample. The dotted lines show the cuts used to measure
    the projected correlation function. \textit{Right:} histograms of
    the $(NUV - r)_{\rm{abs}}$ color, with the same cuts shown on the
    left panel; solid line shows MIS/SDSS sample, and dashed line AIS
    UV-selected sample.}
  \label{fig_mis_nuv_r_sel}
\end{figure}
\input{tab1}

\begin{figure}[!t]%[htbp]
  \epsscale{1.}
  \plotone{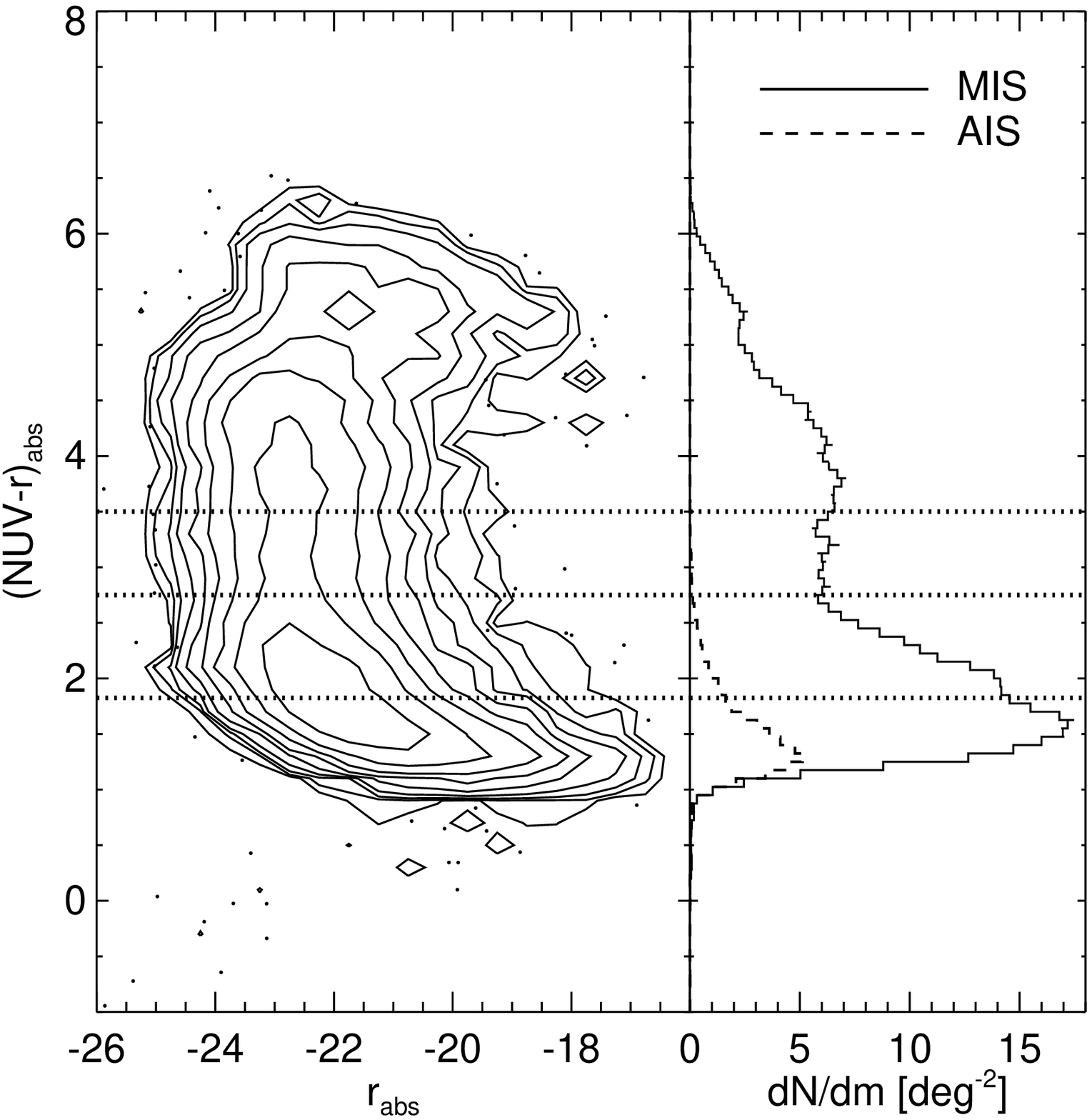}
  \caption{\small Same as fig.\ \ref{fig_mis_nuv_r_sel} but using dust
  extinction corrected absolute magnitudes.}
  \label{fig_mis_nuv_r_sel_dust_cor}
\end{figure}
\input{tab2}

We first consider the MIS sample in order to use the very good
separation between red and blue sequence galaxies in $(NUV -
r)_{\rm{abs}}$ color. The distribution of the galaxies in the
UV-optical color diagram is shown fig.\ \ref{fig_mis_nuv_r_sel}.  As
noticed by previous studies \citep{Martin_2007, Wyder_2007}, the color
distribution shows a clear bimodality between the blue and red clouds
and a transition population located in the so-called green valley. The
distribution of the galaxies in the UV-optical color diagram after
applying the dust extinction correction is shown in
fig. \ref{fig_mis_nuv_r_sel_dust_cor}. The shape of the distribution
is quite different once this correction is applied, with a less
prominent red sequence.

We divided our sample into four bins of $(NUV -r)_{\rm{abs}}$ color in
the following way. We set up first the green valley for both versions
of the color magnitude diagram ($3.75<(NUV - r)_{\rm{abs}} <4.5$
without extinction correction; $2.75<(NUV - r)_{\rm{abs}} <3.5$ with
extinction correction), following previous studies \citep[see fig. 9
and 25 of][]{Wyder_2007}. This also defines a natural cut for the red
and the blue populations. We further divided the blue cloud by the
median color of this population. The cuts we adopted are shown as
dotted lines on figures \ref{fig_mis_nuv_r_sel} and
\ref{fig_mis_nuv_r_sel_dust_cor}.

\begin{figure}[!t]%[htbp]
  \epsscale{1.}
  \plotone{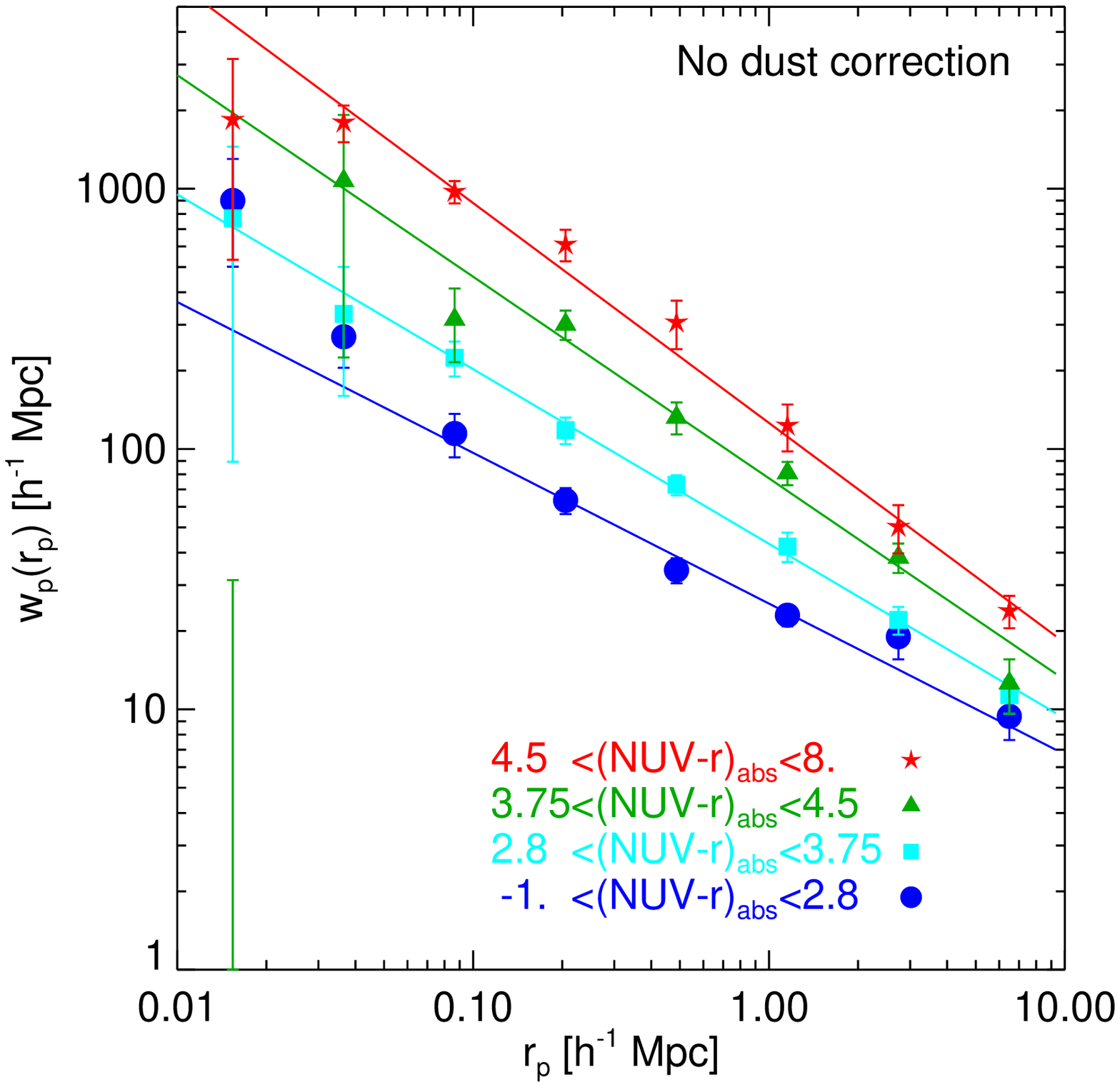}
  \caption{\small Projected correlation functions of MIS/SDSS samples
    cut by $(NUV - r)_{\rm{abs}}$ color (without dust
    extinction correction).}
  \label{fig_mis_nuv_r_wrp}
\end{figure}

\begin{figure}[!t]%[htbp]
  \epsscale{1.}
  \plotone{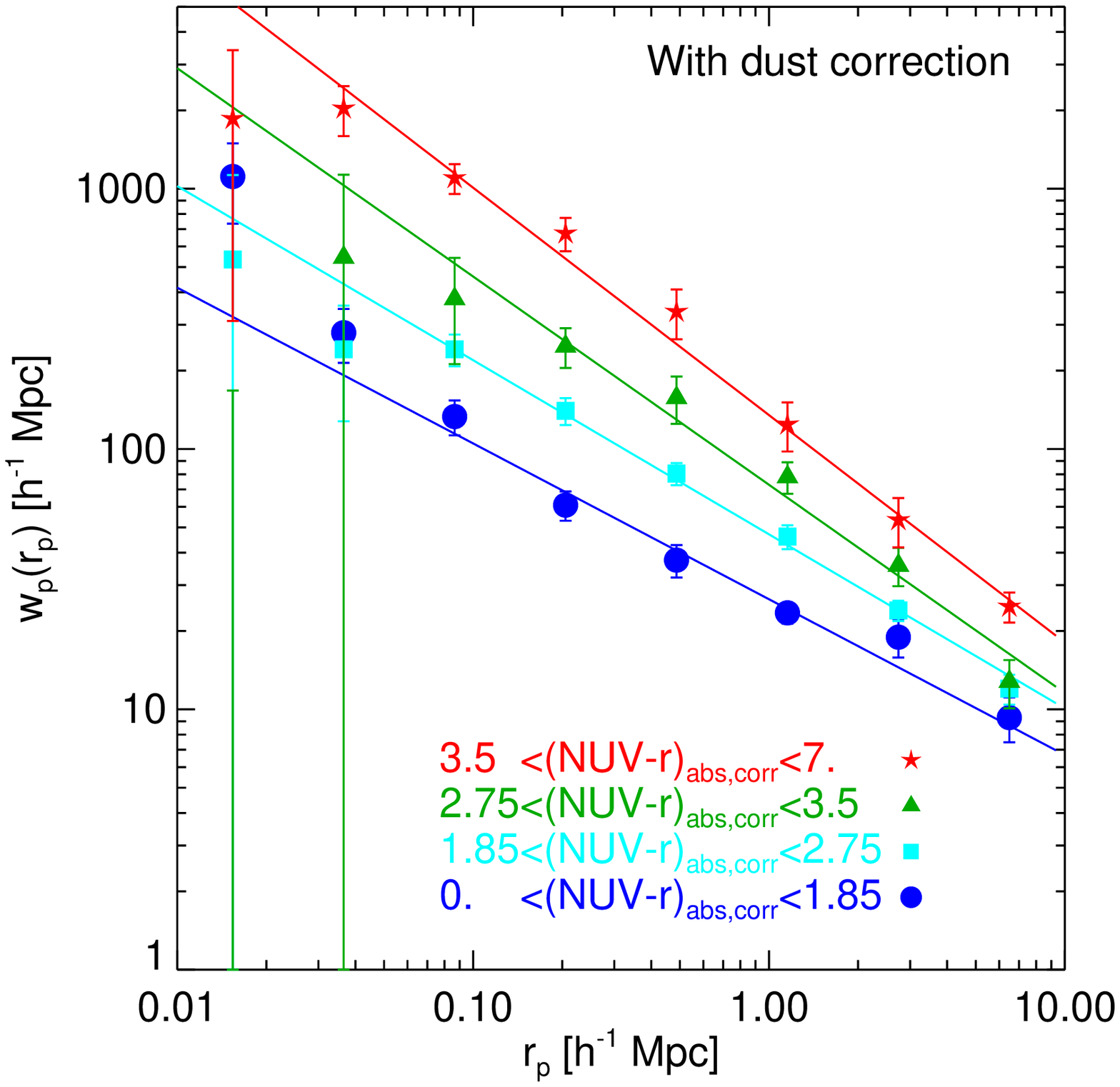}
  \caption{\small Projected correlation functions of MIS/SDSS sample
    cut by $(NUV - r)_{\rm{abs}}$ color (with dust extinction
    correction).}
  \label{fig_mis_nuv_r_wrp_dust_cor}
\end{figure}

The correlation functions are shown fig.\ \ref{fig_mis_nuv_r_wrp} (for
color cuts without dust extinction correction) and fig.\
\ref{fig_mis_nuv_r_wrp_dust_cor} (for color cuts with dust extinction
correction), as well as the power-law fits. In both cases, there is a
monotonic increase of the clustering amplitude for redder galaxies, as
well as a steepening of the correlation function. The results obtained
with (see clustering parameters table \ref{tab_mis_dust_cor}) or
without the dust extinction (table \ref{tab_mis}) correction are very
similar, while the only criterion to define the color cuts has been to
isolate the green valley. This is expected to some extent as the
extinction correction we applied is a function of the uncorrected
$(NUV - r)_{\rm{abs}}$ color and $D(4000)$, which are themselves
fairly well correlated with one another.  \citep[e.g.][]{Martin_2007,
  Wyder_2007}. Note also that the dust corrected green valley overlaps
with the non dust corrected one at the 50\% level. One could argue
that the intermediate clustering of the green valley galaxies we
select with non-dust corrected colors is due to a mix of dusty star
forming galaxies and galaxies with older stellar populations. However,
the fact that this result holds when using dust corrected colors
suggests that dust does not induce such a large population mix in this
region of the color-magnitude diagram. Another possibility is that the
green valley is actually an intrinsic composite population;
\citet{Salim_2007} showed that it is built from normal star forming
galaxies, star forming/AGN composite, and pure AGNs. However, tests
show that within the green valley the large-scale clustering of AGNs
and non AGNs is very similar (see sec.\ \ref{sec_discussion_gv}).

\begin{figure}[!t]%[htbp]
  \plotone{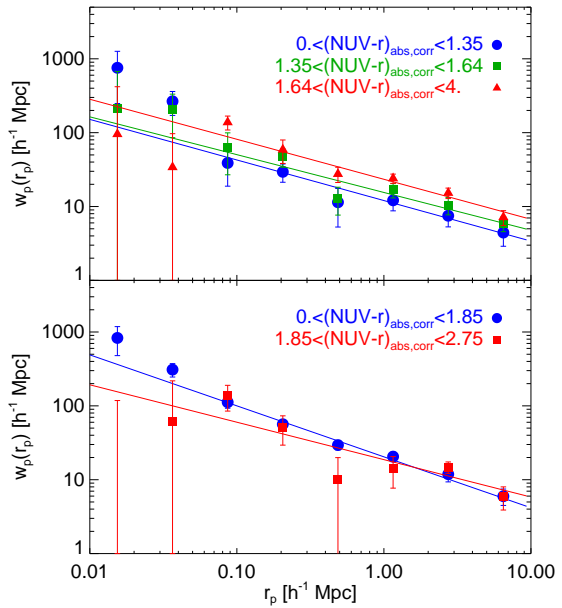}
  \caption{\small Projected correlation functions of AIS UV-selected
    samples in bins of $(NUV - r)_{\rm{abs}}$ color.}
  \label{fig_wrp_ais_nuv_r}
\end{figure}

We also used the AIS sample to investigate the clustering as a
function of the $(NUV - r)_{\rm{abs}}$ color within an UV-selected
sample. The color distribution for this sample is shown on the right
panels of figures \ref{fig_mis_nuv_r_sel} and
\ref{fig_mis_nuv_r_sel_dust_cor} with dashed lines. We used the same
color cuts as for the MIS/SDSS sample and also define three cuts in
order to get similar subsample sizes; we use here colors corrected
from dust extinction. The correlation functions are presented in fig.\
\ref{fig_wrp_ais_nuv_r} and the clustering parameters in table
\ref{tab_ais}. As observed in our MIS/SDSS sample, the correlation
length increases for redder galaxies, although our UV-selected sample
is built mainly from galaxies on the blue sequence (see \S
\ref{sec_data_ais}). This increase is not observed if we use the same
cuts as for the MIS/SDSS sample. The AIS correlation function for the
$1.85<(NUV - r)_{\rm{abs}}<2.75$ bin is noisy because of small
statistics.

Interestingly, our results are poorly fitted by a simple power-law. In
particular at scales $r_p \lesssim 0.1 h^{-1}$ Mpc, the correlation
functions for the blue galaxies in both the MIS and AIS samples show
an excess of pairs, while on the contrary the correlation function of
our reddest bin shows a lack of satellites (figs.\
\ref{fig_mis_nuv_r_wrp} and \ref{fig_mis_nuv_r_wrp_dust_cor}). We note
that for the red galaxies, we are biased against low masses and low specific
star formation rates objects (see fig.\ \ref{fig_m_ssfr}). These faint
red galaxies are known to exhibit a strong clustering at small scales
\citep{Zehavi_2005}. For the blue galaxies, our results suggest a
higher satellite fraction for bluer UV galaxies, which is similar to
what has been observed at high redshift from UV-selected samples
\citep{Lee_2006,Ouchi_2005}.

\input{tab3}

\begin{figure}[!t]%[htbp]
  \plotone{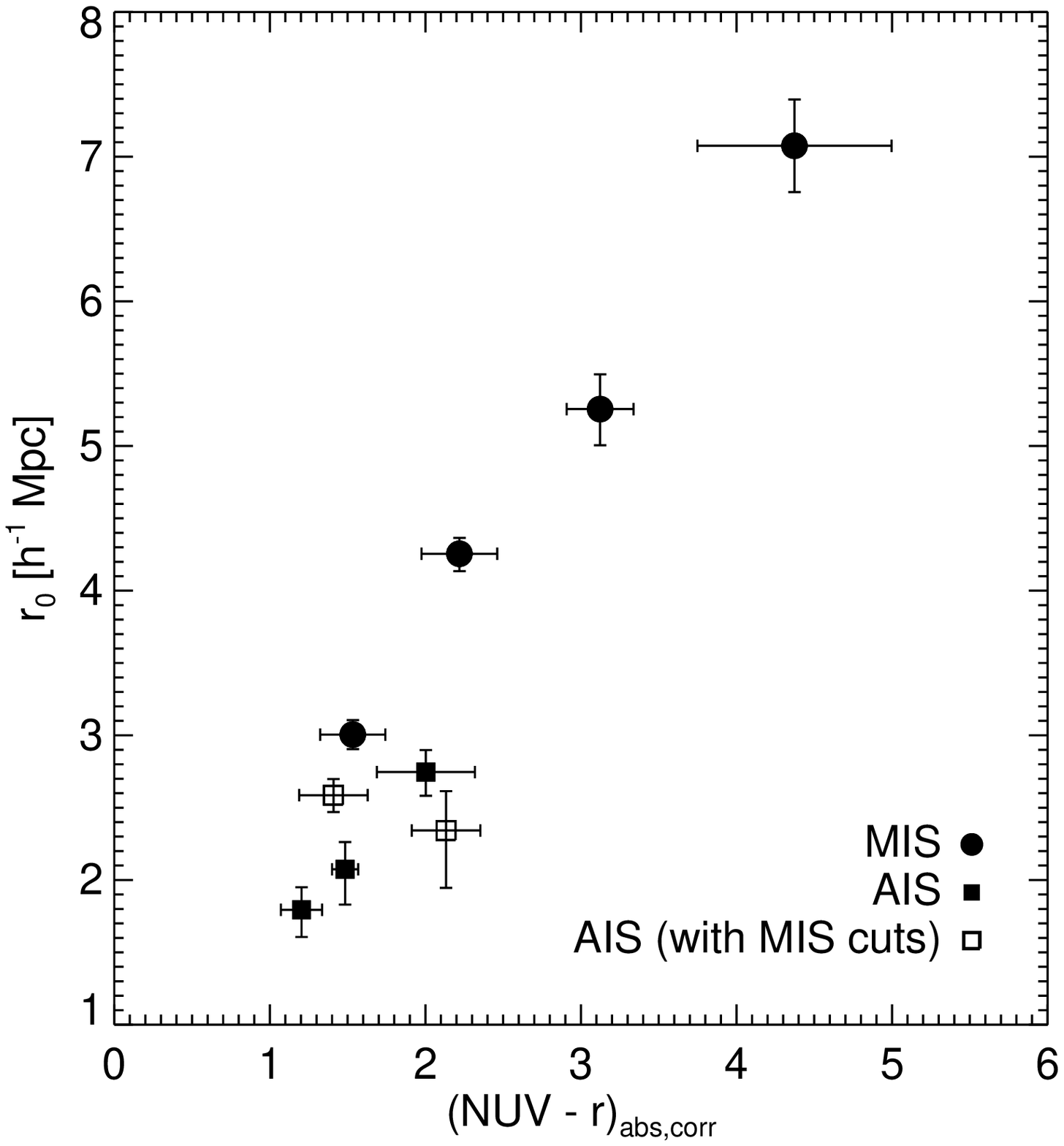}
  \caption{\small Dependence of the correlation length $r_0$ on the
    $(NUV - r)_{\rm{abs, corr}}$ color. The error bars on color are
    the standard deviation of the colors in each bin. Filled circles
    show the MIS/SDSS sample results. Squares show the AIS
    (UV-selected sample) results: open squares using the same cuts as
    for the MIS/SDSS sample, and filled squares cuts defined in order
    to have similar number of galaxies per subsample.}
  \label{fig_ais_mis_ro_nuv_r}
\end{figure}

We show in fig.\ \ref{fig_ais_mis_ro_nuv_r} the combination of our
results for both the AIS (filled and open squares) and MIS samples (filled
circles). Thanks to the cut in $(NUV - r)_{\rm{abs}}$ color, we are
able to probe a large dynamic range in terms of clustering strengths,
with values of the correlation length $r_0$ increasing from $\sim 1.8
h^{-1}$Mpc for the bluest galaxies, to $\sim 7.1 h^{-1}$ Mpc for the
reddest. The UV-selected and MIS/SDSS samples follow the same trends.
However the purely UV-selected galaxies from the AIS survey show a
weaker clustering strength at a given color. The correlation length
from the AIS sample is lower than the one obtained from the MIS/SDSS
sample if the same cuts are applied. The mean color for the AIS
subsample with $0<(NUV - r)_{\rm{abs}}<1.85$ is lower than the
corresponding one for the MIS, but the conclusion holds if different cuts
are applied and the mean color is similar.

\subsection{Comparison with previous studies}
To assess the impact of the NUV selection on our MIS/SDSS sample, we
can directly compare our results with those of \citet{Zehavi_2002},
who divided their sample between blue and red galaxies using the limit
$(u-r)_{abs}$=1.8. They found for red galaxies $r_0 = 6.78 \pm
0.23h^{-1}$ Mpc and $\gamma = 1.86 \pm 0.03$; for blue galaxies $r_0 =
4.02 \pm 0.25h^{-1}$Mpc, $\gamma = 1.41 \pm 0.04$. Applying the same
cuts to our MIS/SDSS sample, we obtain for red galaxies $r_0 = 5.20
\pm 0.15 h^{-1}$ Mpc, $\gamma = 1.77 \pm 0.02$; for blue galaxies $r_0
= 3.24 \pm 0.17 h^{-1}$ Mpc, $\gamma = 1.66 \pm 0.07$. We find the
same trends as \citet{Zehavi_2002} with lower correlation lengths.

We observe a steepening of the slope from $\gamma = 1.58 \pm 0.05$ for
the bluest galaxies to $\gamma = 1.84 \pm 0.03$ for the reddest
for the results from the MIS/SDSS sample without the dust
extinction correction. This increase of the slope with color is
similar to what has been already observed by previous studies from
optical samples at low redshift. Selecting active and passive galaxies
from their spectra, \citet{Madgwick_2003} found that they exhibit
power laws with respective slopes $\gamma = 1.60 \pm 0.04$ and $\gamma
= 1.95 \pm 0.03$. \citet{Budavari_2003} observed similar trends in their
angular correlation function measurements as a function of spectral type. 
Their blue samples have a mean slope $\gamma = 1.67 \pm
0.11$ and their red samples have $\gamma = 1.99 \pm 0.06$.

The correlation length we obtained from the MIS/SDSS sample for the
bluest galaxies in $(NUV - r)_{\rm{abs}}$ of $r_0 \sim 3.0 \pm
0.1h^{-1}$Mpc is lower than the values of $3.6-4.5 h^{-1}$Mpc found
for blue galaxies in optical samples such as the SDSS
\citep{Budavari_2003, Zehavi_2005} or 2dF \citep{Madgwick_2003,
  Norberg_2002} at similar redshifts. This lower value is due to the
combination of our selection which biases against low stellar mass and
low specific star formation rate objects, and the larger star
formation history dynamic range probed by the $(NUV - r)_{\rm{abs}}$
color as well.

The $r_0$ value we measured for the bluest bin of our UV-selected
sample, $r_0 = 1.79 \pm 0.17h^{-1}$Mpc, is among the lowest observed
in the local Universe. It is slightly lower than the value $r_0 =
2.31^{+0.5}_{-0.4} h^{-1}$Mpc we obtained from a flux-limited
(NUV$<$21.) UV-selected sample \citep{Milliard_2007}.

%This estimate is lower than the ones obtained for blue galaxies from
%SDSS \citep{Budavari_2003, Zehavi_2005} or 2dF data
%\citep{Madgwick_2003, Norberg_2002} at similar redshifts, whose
%measured values of $r_0$ range between $3.6 h^{-1}$Mpc and
%$4.5h^{-1}$Mpc.

These estimates are closer to the values derived for gas-rich, H-I
selected galaxies obtained by \citet{Meyer_2007} of $r_0 = 2.70\pm
0.21 h^{-1}$Mpc and $\gamma = 1.56\pm0.1$. This correspondence might
indicate that we can associate the bluest galaxies with gas-rich
galaxies, which is in agreement with the fact that the gas fraction of
blue galaxies is larger \citep[see, e.g.,][]{Schiminovich_2007}.

Transition galaxies in the green valley indeed exhibit an intermediate
clustering pattern, although comparison with other studies might lead
to a different conclusion depending on the selections used. For
instance the correlation length we measure is close to that derived by
\citet{Zehavi_2005} for red galaxies selected by $g-r$ color and
$-20<\rm{r}_{\rm{abs}}<-19$ ($r_0= 5.7 h^{-1}$ Mpc). On the other
hand, our $r_0$ value for the green valley is also similar to one of
the faint samples of blue galaxies studied by \citet{Budavari_2003}
(T3 sample, $r_0= 5.9 \pm 0.51 h^{-1}$ Mpc), who selected galaxies
according to their spectral energy distribution. At $z\sim 0.85$
\citet{Coil_2008} measured the projected correlation function of green
galaxies, selected using the $(U-B)$ restframe color. The correlation
length of $r_0 = 5.17 \pm 0.42 h^{-1}$Mpc that they derived is similar
to the one we obtained.

\section{Discussion}\label{sec_discussion}

\subsection{Intermediate clustering strength of green valley galaxies}\label{sec_discussion_gv}
\begin{figure}[!t]%[htbp]
  \plotone{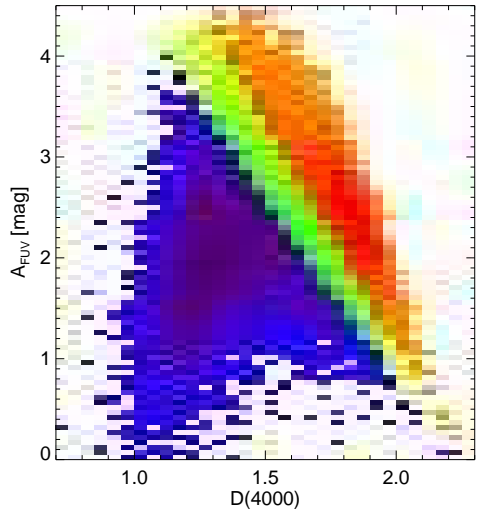}
  %\plotone{plot_MIS_afuv_d4000_mw_all_whole_sample_cut23.eps}
  \caption{\small Extinction correction in the FUV band, $A_{FUV}$, as
    a function of the 4000 \AA~break, D(4000) for the MIS/SDSS
    sample. The extinction correction is obtained from eq.\
    \ref{eq_dust_cor}. The three color shades code objects on the blue
    sequence, green valley and red sequence, from left to right.}
  \label{fig_mis_afuv_d4000}
\end{figure}

The results presented in sect.\ \ref{sec_results} show that green
valley galaxies have an intermediate clustering strength compared to
bluer or redder galaxies. This holds whether dust extinction is
applied or not. As the green valley is a mix of intermediate star
formation history galaxies and dusty star forming ones, interpretation
of this result relies on the efficiency of the dust correction
procedure to distinguish between these populations. Figure
\ref{fig_mis_afuv_d4000} shows the extinction correction in the FUV
band, $A_{FUV}$ as a function of the 4000 \AA~break, D(4000) for the
MIS/SDSS sample. Galaxies within the green valley ($3.75<(NUV -
r)_{\rm{abs}} <4.5$) lie in the middle (green) region of this
plot. This figure shows that the dust correction prescription we used
takes into account the evolutionary stage of the galaxy
population. For instance, more evolved galaxies with high D(4000) or
low specific star formation rates have a lower extinction correction
than younger (low D4000) or more active galaxies (high specific star
formation rates), in agreement with recent analysis from
\citet{Cortese_2008}. While still subject to some uncertainties, this
procedure should clean up the green valley population from the highly
dusty population.

%Green valley galaxies with older stellar populations (higher
%D(4000)) have a lower dust correction.  This suggests that the
%intermediate clustering of the green valley galaxies we observe is not
%due to a mix of dusty star forming galaxies and galaxies hosting older
%stellar populations.

\citet{Martin_2007} noted that the fraction of galaxies with AGNs (in
their case type 2, or narrow-line AGNs) peaks in the green
valley. There is no study devoted to the clustering of type 2 AGNs to
which we can directly compare our results. \citet{Li_2006b} studied
however the cross-correlation of narrow-line AGNs with galaxies within
SDSS. In particular, they showed that their clustering is in agreement
with the assumption that a high fraction of them are within central
galaxies in dark matter halos, and that they reside mainly within
halos with masses $10^{12}$-$10^{13} M_{\odot}$, values consistent
with our results for the green valley (see \S
\ref{sec_discussion_dmh_mass}). Note also that several studies pointed
out the ``intermediate'' nature of galaxies hosting AGNs
\citep{Heckman_2004, Kauffmann_2003b}. For instance, the average
spectra of AGNs galaxies host is remarkably similar to those of disk
galaxies with Sb Hubble type \citep{Yip_2004a, Yip_2004b}.

The intermediate clustering we measured for transition galaxies may be
then mainly driven by galaxies hosting AGNs. To test this further, we
restricted our MIS/SDSS sample to the overlap with SDSS DR4, in order
to discriminate, within the green valley, AGNs and non AGNs, based on
the catalogs from \citet{Kauffmann_2003c}. \citet{Kauffmann_2003c}
selected type 2 AGNs from star forming galaxies using the diagram
proposed by \citet{Baldwin_1981}. We note that the objects classified
as 'AGNs' by \citet{Kauffmann_2003c} include star forming/AGNs
composites; the low numbers of this sample do not enable to perform
clustering measurement separately on these populations.

\begin{figure}[!t]%[htbp]
  \plotone{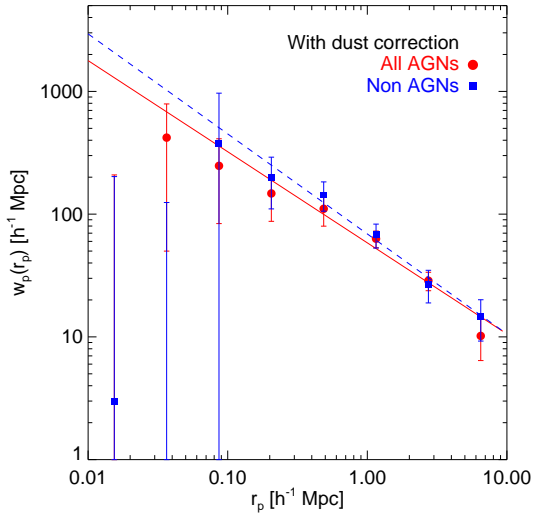}
  \caption{\small Projected correlation functions for MIS green valley
    galaxies, assuming dust extinction correction. The sample is here
    restricted to the overlap with SDSS DR4, and galaxies divided
    between galaxies with (filled circles) or without AGNs (filled
    squares).}
  \label{fig_wrp_green_valley_agn}
\end{figure}

We computed the clustering of galaxies within the green valley as
defined in sect.\ \ref{sec_results} and using dust correction, but
separately for galaxies with or without AGNs. The AGNs represent 56\%
of the green valley. The projected correlation functions are presented
in fig.\ \ref{fig_wrp_green_valley_agn}. The main differences in terms
of clustering between these two population, as noted already by
\citet{Li_2006b}, are at small scales. At scales $0.1<r_p<0.5 h^{-1}$
Mpc, the non-AGNs galaxies exhibit marginally more pairs than the
AGNs. At large scales, the clustering of AGNs and non-AGNs populations
is very similar; the correlation length for the AGNs is $4.7 \pm
0.3h^{-1}$Mpc, and the correlation length for the non-AGNs is
$5.1\pm0.4 h^{-1}$ Mpc. This suggests that the intermediate clustering
of the green valley galaxies is genuine and not linked to the
predominance of AGNs in this region of the UV optical color-magnitude
diagram.

\subsection{Dark Matter Halo mass}\label{sec_discussion_dmh_mass}
We used the analytical models of \citet{Mo_2002} to estimate the
minimum mass of the dark matter halos that have similar clustering, by
matching the bias parameter $b_8$, computed at 8 $h^{-1}$Mpc,
following e.g. \citet{Magliocchetti_2000}. This comparison shows that
the mass of the host dark matter halos increases smoothly with $(NUV -
r)_{\rm{abs}}$. Blue galaxies have clustering similar to halos with
minimum masses around $10^{11} h^{-1}M_{\odot}$, green valley galaxies
$10^{12.5} h^{-1} M_{\odot}$, and red galaxies $10^{13} h^{-1}
M_{\odot}$. The minimum halo mass for blue galaxies is in agreement
with estimates derived from UV-selected galaxies at low redshift
\citep{Heinis_2007, Milliard_2007}, and is additional evidence that
active star formation preferentially occurs in low density
environments at low redshifts \citep{Gomez_2003, Lewis_2002}.

The minimum dark halo mass for transition galaxies is similar to those
of galaxy groups. This suggests that galaxies within the green valley
mainly reside in groups. At $z\sim 0.85$ green valley galaxies have a
higher bias, $b = 1.65\pm 0.26$ \citep{Coil_2008}; at these redshifts
this corresponds also to halos with minimum masses of the order of
$10^{12.5} h^{-1} M_{\odot}$. The combination of these results show
that green valley galaxies reside in groups of similar masses at
$z\sim 0.85$ and $z = 0$. We note also that within the green valley
the correlation function of star forming galaxies is marginally larger
at small scales than the AGN correlation function (see fig.\
\ref{fig_wrp_green_valley_agn}). This could provide some hints about
the nature of those objects in the context of the Halo Occupation
Distribution and the satellite versus central galaxy population. As
AGNs are expected to be preferentially central galaxies
\citep{Li_2006b} this is a hint that transition galaxies could be
mainly satellites rather than central galaxies within groups and may
preferentially lie in the outskirts of their host DMHs as a recently
accreted population, as proposed by \citet{Coil_2008}. We can
speculate on the nature of the mechanisms besides AGN feedback
involved in the quenching of star formation in transition galaxies, if
they are related to their environment. In this case, given the low
relative velocities in galaxy groups, local processes such as tidal
mechanisms may induce significant suppression of star formation
\citep[e.g.][]{Boselli_2006, Bower_2004, Gomez_2003, Lewis_2002}.

Finally, as expected, the reddest galaxies have a clustering strength
similar to large groups or clusters.

These results show that there is a correlation between star formation
history and dark matter halo mass, as the galaxies having formed stars
more recently inhabit less massive halos. These findings are in
agreement with the ``downsizing'' scenario \citep{Cowie_1996}, which
states that more massive galaxies formed their stars at higher
redshifts \citep[see e.g.][]{Bundy_2006, DeLucia_2006}.

\subsection{Halo occupation limits}\label{MIS_hon}
\begin{figure}[htbp]%[!t]
  \plotone{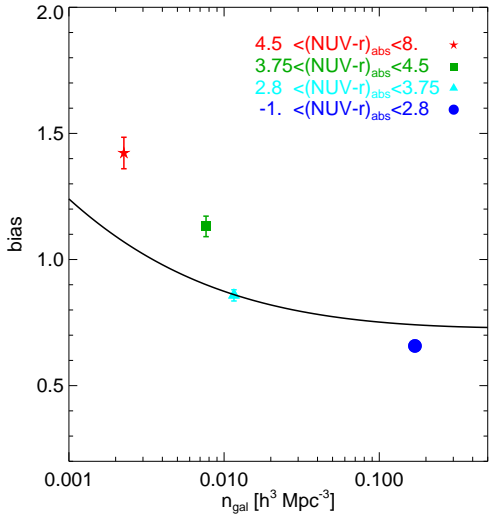}
  \caption{\small Bias as a function of galaxy number density for
    $(NUV - r)_{\rm{abs}}$ cut samples (same color legend as fig.\
    \ref{fig_mis_nuv_r_wrp}). We show here only results without the
    dust extinction correction. The solid line shows the relation
    expected for dark matter halos, using \citet{Mo_2002} models.}
  \label{fig_mis_b_n}
\end{figure}

Our approach in the previous section to estimate the dark matter halo
mass from the clustering with \citet{Mo_2002} models has the caveat
that it assumes only one galaxy per halo. This assumption is quite
unlikely according to recent numerical and observational studies
\citep[e.g.,][]{Kravtsov_2004,Ouchi_2005}. In the framework of the
Halo Occupation Distribution (HOD) \citep[e.g.,][]{Cooray_2002}, the
number of galaxies in a given halo depends on its mass. We can however
use \citet{Mo_2002} models to get a first estimate of the dependence
of the halo occupation with star formation history. In fig.\
\ref{fig_mis_b_n} we compare the bias and the number density of the
samples with those of the dark matter halos. We computed the number
densities of the various samples using the luminosity functions
derived by \citet{Wyder_2007} in $(NUV - r)_{\rm{abs}}$ bins from a
similar MIS sample. These luminosity functions are not available for
dust extinction corrected samples, so we only show on fig.\
\ref{fig_mis_b_n} our MIS results obtained without any dust
correction. As the clustering results are in very good agreement with
or without applying the dust correction, and as we want to get only
some qualitative limits on the halo occupation numbers, we expect our
conclusions to be robust.

The relation expected for dark matter halos from the \citet{Mo_2002}
models is shown by the solid line. On this plot, the dark matter halo
mass increases on this line from right to left. As this model assumes
that each halo hosts one galaxy, measurements on the solid line
indicate an average halo occupation number of 1, measurements above an
average halo occupation greater than 1, and measurements below an
average occupation lower than 1. Our measurements recover the trend
expected for dark matter halos, as galaxy populations with lower
number densities are more clustered. However, galaxies do not lie on
the dark matter halos bias-density relation, which is due to the fact
that the halo occupation varies among populations. Our results show
that the average occupation number increases with $(NUV -
r)_{\rm{abs}}$ color. Bluest galaxies in the local Universe have a
very low occupation number in halos that have similar clustering
strength. Our second blue sample contains mainly galaxies with an
average occupation number close to 1. Transition and red galaxies have
an occupation number around 5. Note that the occupation number for the
red galaxies is potentially an under estimation given that our
selection biases against low mass and low specific star formation
rates galaxies (see sect. \ref{sec_res_nuv_r})

These results are in qualitative agreements with numeric
\citep{Berlind_2003, Zheng_2005}, as well as observational studies
\citep{Hogg_2003, Magliocchetti_2003, Zehavi_2005}.

\subsection{Specific star formation rate vs. stellar mass}\label{sec_ssfr}

\begin{figure}[!t]%[htbp]
  \plotone{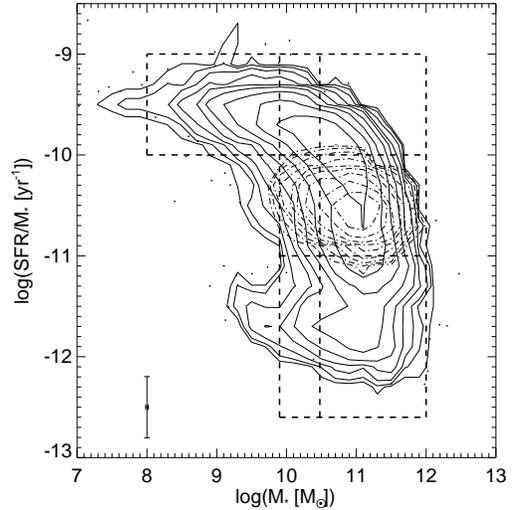}
  \caption{\small Specific star formation rate as a function of
  stellar mass. The dot-dashed contours show the distribution of the
  green valley defined with $ 2.75 < (NUV - r)_{\rm{abs,corr}} <
  3.5$. The dashed lines represent the cuts we consider for clustering
  measurements. The error bars show the dispersion caused by the dust
  extinction correction.}
  \label{fig_m_ssfr}
\end{figure}

The results we present above add further evidence that the clustering
of galaxies strongly depends on star formation history. However galaxy
clustering also depends on stellar mass \citep[e.g.][]{Li_2006a,
  Meneux_2008}; we investigate which of these two parameters shows the
strongest correlation with environment. To do so, we consider an
alternative to the UV-optical color-magnitude diagram, by plotting the
specific star formation rate $SFR/M_{\ast}$ as a function of the
stellar mass $M_{\ast}$ in figure \ref{fig_m_ssfr}.

We computed the star formation rates from the NUV luminosities
corrected for dust extinction (see \S \ref{sec_dust_cor}) using the
\citet{Kennicutt_1998} relation scaled to a \citet{Kroupa_2001} initial
mass function:

\begin{equation}
  \textrm{SFR }(M_{\odot} \textrm{yr}^{-1})= 10^{-28.02}L_{\nu} (\textrm{erg s$^{-1}$ Hz$^{-1}$}) \textrm{.}
\end{equation}

We estimate the stellar masses using the equation derived by
\citet{Yang_2007} from the relation between stellar mass-to-light
ratio and color of \citet{Bell_2003}:

\begin{eqnarray}
  \log \left[\frac{M_{\ast}}{M_{\odot}} \right] & = & -0.306 + 1.097[(g_{abs} - r_{abs})_{corr}] \nonumber\\ 
  & & - 0.1 - 0.4[r_{abs,corr} -4.64]
\end{eqnarray}

which implicitly assumes a \citet{Kroupa_2001} initial mass function.

\begin{figure}[!t]%[htbp]
  \plotone{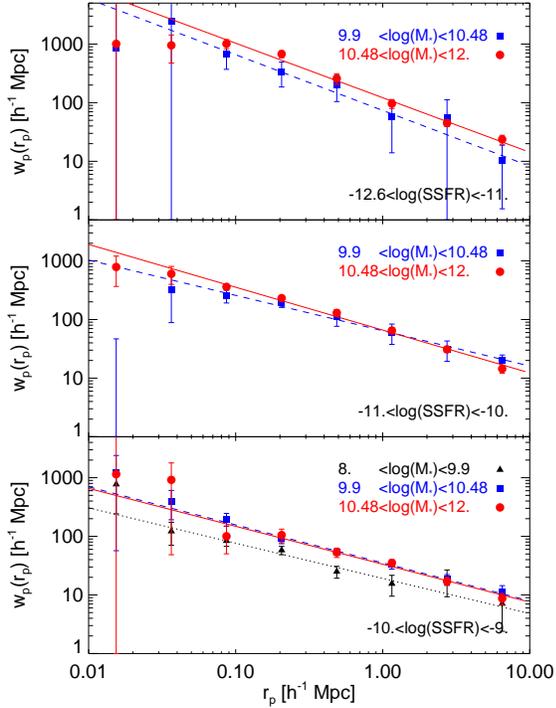}
  \caption{\small Projected correlation functions of the MIS/SDSS
    samples cut by stellar mass and specific star formation rate. The
    top panel shows measurements obtained for the
    $-12.6<\log(SSFR)<-11$ bin, the middle panel for the
    $-11<\log(SSFR)<-10$ bin, and the bottom panel for the
    $-10<\log(SSFR)<-9$ bin. On each panel, filled squares show
    results from $9.9<\log(M_{\ast})<10.48$, filled circles from
    $10.48<\log(M_{\ast})<12$, and filled triangles
    $8<\log(M_{\ast})<9.9$ (only for $-10<\log(SSFR)<-9$).}
  \label{fig_wrp_m_ssfr}
\end{figure}
\input{tab4}

The cuts we adopted are shown as dashed lines in fig.\
\ref{fig_m_ssfr}. Within the stellar mass range
$10^{9.9}<M_{\ast}<10^{12} M_{\odot}$, we choose an additional mass
cut at $M_{\ast} = 3\times 10^{10} M_{\odot}$. Several studies indeed
showed that in the local Universe this value divides galaxies between
active or passive in terms of star formation
\citep{Kauffmann_2003a}. Moreover, according to numerical studies
\citep{Keres_2005, Keres_2008}, the way galaxies accrete gas depends
on stellar mass. Galaxies below $3\times 10^{10} M_{\odot}$ are
dominated by a ``cold mode'' of gas accretion, while galaxies above
this mass are dominated by a ``hot mode'' of gas accretion. The
geometry of these two modes are also different, the cold mode being
mostly filamentary, while the hot mode quasi-spherical, hence implying
a link with environment. We then further divide our subsample in mass
by this value $M_{\ast} = 3\times 10^{10} M_{\odot}$, and also use two
cuts in specific star formation rate at $\log(SFR/M_{\ast} [yr^{-1}])
= 10^{-11}$ and $\log(SFR/M_{\ast} [yr^{-1}]) = 10^{-10}$.

The correlation functions are shown fig.\ \ref{fig_wrp_m_ssfr}. Each
panel of fig.\ \ref{fig_wrp_m_ssfr} shows a given cut in specific star
formation rate, and the relevant cuts in stellar mass. While there are
some differences in the small-scale clustering ($r_p \lesssim 0.5
h^{-1}$ Mpc), at larger scales the correlation functions for the
galaxies with $10^{9.9}<M_{\ast}<10^{12} M_{\odot}$ are remarkably
similar for a given specific star formation rate. The main difference
we observe in clustering strength is for the less massive sample
($10^{8.}  <M_{\ast}< 10^{9.9} M_{\odot}$) within the highest specific
star formation rate bin. There is also a difference for the lowest
specific star formation rate bin, but in this bin the low mass sample
with $10^{9.9}<M_{\ast}<10^{10.48} M_{\odot}$ has very low statistics
so its correlation function must be interpreted with caution.

\begin{figure}[!t]%[htbp]
  \plotone{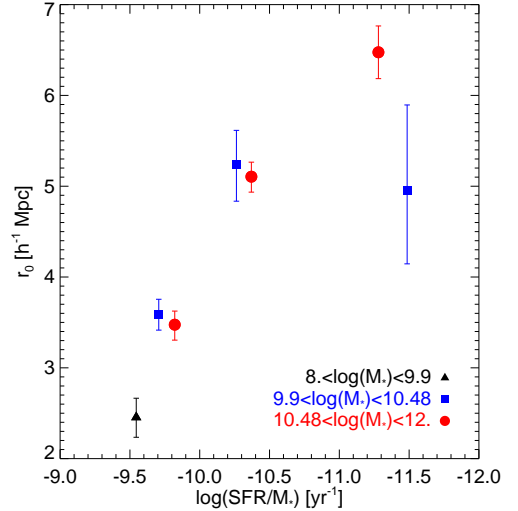}
  \caption{\small Correlation length as a function of specific star
    formation rate and stellar mass. The filled triangle shows the
    result from $8.<\log(M_{\ast})<9.9$, filled squares show results
    from $9.9<\log(M_{\ast})<10.48$, and filled circles from
    $10.48<\log(M_{\ast})<12.$.}
  \label{fig_ro_m_ssfr}
\end{figure}

We show in fig.\ \ref{fig_ro_m_ssfr} the correlation lengths derived
from these measurements as a function of specific star formation rate
and stellar mass. The global trend is that $r_0$ decreases with
specific star formation rate. For $-11<\log(SFR/M_{\ast})<-9$ and
$9.9<log(M_{\ast})<12$, the correlation length does not depend on
stellar mass in a given specific star formation rate range. This does
not seem to be true in our extreme stellar mass bins. However the
statistics of our $-12.6<\log(SFR/M_{\ast})<-11 ,\quad
9.9<\log(M_{\ast})<10.48$ sample are small; and because of our
selection, we can not compare to lower specific star formation rates
at $8.<\log(M_{\ast})<9.9$. Nevertheless, these results suggest that
the specific star formation rate shows a stronger correlation with
environment (as traced by the large-scale clustering) than the stellar
mass. Interestingly, this is true below and above the transition mass
of $3\times 10^{10} M_{\odot}$. Within the ranges
$-11<\log(SFR/M_{\ast})<-9$ and $9.9<\log(M_{\ast})<12$, our results
show that there is a smooth transition of the large-scale clustering
strength, and hence in dark matter host halo masses, as a function of
star formation history, independently of stellar mass. This result is
consistent with our measurements based on color selected samples
(because of the correlation between specific star formation rate and
$(NUV-r)_{\rm{abs}}$), but provides a direct interpretation of the
dependence of clustering on galaxies physical parameters (stellar mass
and specific star formation rate).

%dependency of clustering with physical quantities (stellar mass and
%specific star formation rate) of the galaxy properties.

%The clustering parameters derived from the power-law fit (listed in
%table \ref{tab_mis_m_ssfr}) suggest that the specific star formation
%rate shows a stronger correlation with environment (as traced by the
%large-scale clustering) than the stellar mass. This is true for
%galaxies with masses $10^{10.48}<M_{\ast}<10^{12.}  M_{\odot}$ over
%the whole specific star formation rate we explore here, and also for
%$10^{9.9}<M_{\ast}<10^{10.48} M_{\odot}$ on a smaller range of
%specific star formation rate ($ 10^{-11}< \log(SFR/M_{\ast} [yr^{-1}])
%< 10^{-9}$), given the statistics of our sample. Within these ranges
%of stellar masses and specific star formation rates, our results show
%that there is a smooth transition of the large-scale clustering
%strength, and hence in dark matter host halo masses, as a function of
%star formation history, independently of stellar mass. 

In the local Universe, high specific star formation rate galaxies are
anti-biased ($b_8 <1$), and have clustering similar to dark matter
halos of minimum masses $\sim 10^{11} M_{\odot}$. Galaxies with
intermediate specific star formation rates and within the green valley
are unbiased ($b_8 \sim 1$), and are hosted by dark matter halos of
minimum masses $\sim 10^{12} M_{\odot}$. Lower specific star formation
rate galaxies are biased ($b_8 > 1$); their clustering strength is
similar to dark matter halos of minimum masses $\sim 10^{13}
M_{\odot}$.

These results are in agreements with those of \citet{Blanton_2005} who
noticed that color ($g-r$ in their case, which is also a tracer of
star formation history) is the galaxy property most predictive of
environment. \citet{Kauffmann_2004} showed that the galaxy property
most sensitive to environment is star formation history (and in
particular the specific star formation rate), and although the
amplitude of the dependence varies with mass, it is observed at all
masses (for $10^{9.5}<M_{\ast}<10^{11} M_{\odot}$).

\section{Conclusions}\label{sec_conclusion}
We used the unique combination of GALEX UV photometry and SDSS
spectroscopy to perform spatial clustering studies from UV and
UV/optical-selected samples. We take advantage of the remarkable
correlation between the $(NUV - r)_{\rm{abs}}$ color with star
formation history to investigate directly how it relates to
large-scale clustering.

The results from the UV-selected samples, in addition to those from
\citet{Heinis_2007} and \citet{Milliard_2007}, confirm the low
clustering of UV-selected galaxies in the local Universe. This is
further evidence that active star formation takes preferentially place
in low density environments at low redshifts \citep{Gomez_2003,
  Lewis_2002}. Within the blue sequence, redder UV-selected galaxies
are more clustered, but the clustering strength of UV-selected
galaxies is however lower than for optically-selected galaxies at a
given $(NUV - r)_{\rm{abs}}$ color.

The results from our MIS/SDSS sample recover the expected trends from
previous studies that redder galaxies are more clustered. Thanks to
the greater sensitivity of the $(NUV - r)_{\rm{abs}}$ color to star
formation history than optical colors, our results span a wide range
in clustering strength.  Our results show that galaxies that
experienced star formation more recently inhabit less massive halos,
which is consistent with the ``downsizing'' scenario
\citep[e.g.][]{Bundy_2006, Cowie_1996, Neistein_2006}. Also, their
average halo occupation is lower, in such a way that the bluest
galaxies are hosted by only a small fraction of the dark matter halos
that have the same clustering. These findings show that, in contrast
to what is observed from LGBs at high redshift \citep[see
e.g.][]{Giavalisco_2001, Lee_2006}, star formation is not primarily
driven by gravitation in the local Universe. The clustering of the
green valley galaxies suggests that they are mainly satellites within
galaxy groups. Comparison with previous work of \citet{Coil_2008}
indicate that these transition galaxies reside in similar environments
at $z\sim 1 $ and $z=0$.

Our results show that there are smooth trends between star formation
history and environment. It is interesting to notice that based on the
results from optical selections, one might not expect to observe such
smooth transitions. Previous studies showed that there is a bimodality
in a number of galaxy properties: color \citep[$u-r$ for instance;
see][]{Baldry_2004, Strateva_2001}, stellar mass
\citep{Kauffmann_2003b}, or clustering
\citep{Budavari_2003}. \citet{Wyder_2007} already noticed however that
the $(NUV - r)_{\rm{abs}}$ color distribution in the local Universe is
not as bimodal as the $u - r$ color (see also our fig.\
\ref{fig_mis_nuv_r_sel}), due to the greater sensitivity of the UV
spectral range to star formation activity, and reddening by
interstellar dust. Our results suggest that in the local Universe,
rather than a bimodality, there is a smooth transition in global
environments (as probed by the clustering) for galaxies with different
star formation histories (as probed by the $(NUV - r)_{\rm{abs}}$
color). These results hold whether dust extinction is applied or not,
and are insensitive to the predominance of AGNs among transition
galaxies.

Finally, our results show that the large-scale clustering trends we
observe are mainly driven by star formation history, and not stellar
mass. This is indication that at a fixed star formation history, there
is no strong variation of the host dark matter halo mass. Our results
suggest that the environmental dependance is first driven by the star
formation history and that the stellar mass only acts as a secondary
parameter.

\acknowledgements 

We thank Ching-Wa Yip for useful discussions.  GALEX (Galaxy Evolution
Explorer) is a NASA Small Explorer, launched in April 2003. We
gratefully acknowledge NASA’s support for construction, operation, and
science analysis for the GALEX mission, developed in cooperation with
the Centre National d’Etudes Spatiales of France and the Korean
Ministry of Science and Technology.

{\it Facilities:} \facility{GALEX}, \facility{Sloan}

\end{document}

%% file: tab1.tex
\begin{deluxetable*}{ccccccc}
\tablecolumns{5} \tabletypesize{\footnotesize} \tablewidth{0pt}
\tablecaption{\small MIS/SDSS samples clustering results\tablenotemark{*}\label{tab_mis}}
\tablehead{
\colhead{Sample} & \colhead{$N_{gal}$} &\colhead{$\langle z \rangle$} & \colhead{$r_0$ [$h^{-1}$ Mpc]} & \colhead{$\gamma$} & \colhead{$b_8$}}
\startdata
$-1.0\phantom{0}<(NUV - r)_{\rm{abs}} <2.8\phantom{0}$ & 6293 & 0.08 & 2.93 $\pm$ 0.09&1.58 $\pm$ 0.05 & 0.66 $\pm$ 0.02\\[0.1cm]
$\phantom{-}2.8\phantom{0}<(NUV- r)_{\rm{abs}} <3.75$ & 6291 & 0.10 & 4.0\phantom{0} $\pm$ 0.12&1.67 $\pm$ 0.03 & 0.86 $\pm$ 0.02\\[0.1cm]
$\phantom{-}3.75<(NUV- r)_{\rm{abs}} <4.5\phantom{0}$ & 2587 & 0.10 & 5.48 $\pm$ 0.21 &1.77 $\pm$ 0.04 & 1.13 $\pm$ 0.04\\[0.1cm]
$\phantom{-}4.5\phantom{0}<(NUV - r)_{\rm{abs}} <8.0\phantom{0}$  & 6702 & 0.10 &6.93 $\pm$ 0.32 &1.84 $\pm$ 0.03 & 1.43 $\pm$ 0.06\\[0.1cm]
\enddata
\tablenotetext{*}{Without dust extinction correction}
\end{deluxetable*}

%% file: tab2.tex
\begin{deluxetable*}{ccccccc}
\tablecolumns{5} \tabletypesize{\footnotesize} \tablewidth{0pt}
\tablecaption{\small MIS/SDSS samples clustering results\tablenotemark{*}\label{tab_mis_dust_cor}}
\tablehead{
\colhead{Sample} & \colhead{$N_{gal}$} &\colhead{$\langle z \rangle$} & \colhead{$r_0$ [$h^{-1}$ Mpc]} & \colhead{$\gamma$} & \colhead{$b_8$}}
\startdata
$0.\phantom{00}<(NUV- r)_{\rm{abs,corr}} <1.85$ & 6635 & 0.08 & 3.0\phantom{0} $\pm$ 0.1\phantom{0}&1.60 $\pm$ 0.05 & 0.67 $\pm$ 0.02\\[0.1cm]
$1.85<(NUV - r)_{\rm{abs,corr}} <2.75$ & 6534 & 0.10 & 4.25 $\pm$ 0.12&1.67 $\pm$ 0.03 & 0.89 $\pm$ 0.02\\[0.1cm]
$2.75<(NUV - r)_{\rm{abs,corr}} <3.5\phantom{0}$ & 2769 & 0.10 & 5.25 $\pm$ 0.24 &1.80 $\pm$ 0.05 & 1.09 $\pm$ 0.05\\[0.1cm]
$3.5\phantom{0}<(NUV - r)_{\rm{abs,corr}} <7.0\phantom{0}$  & 5816 & 0.10 &7.07 $\pm$ 0.32 &1.87 $\pm$ 0.03 & 1.46 $\pm$ 0.06\\[0.1cm]
\enddata
\tablenotetext{*}{With dust extinction correction}
\end{deluxetable*}

%% file: tab3.tex
\begin{deluxetable*}{cccccc}
\tablecolumns{5} \tabletypesize{\footnotesize} \tablewidth{0pt}
\tablecaption{\small UV-selected samples clustering results\label{tab_ais}}
\tablehead{
\colhead{Sample} & \colhead{$N_{gal}$} & \colhead{$\langle z \rangle$} & \colhead{$r_0$[$h^{-1}$ Mpc]} & \colhead{$\gamma$} & \colhead{$b_8$}}
\startdata
$0.\phantom{00}<(NUV - r)_{\rm{abs,corr}} <1.85$& 10861 & 0.05 &2.59$\pm$0.11  & 1.69 $\pm$ 0.05 & 0.57 $\pm$ 0.02\\[0.1cm]
$1.85<(NUV - r)_{\rm{abs,corr}} <2.75$ &\phantom{0}2596 & 0.06 & 2.34$^{+0.27}_{-0.40}$  & 1.51 $\pm$ 0.12 & 0.56$^{+0.05}_{-0.07}$\\[0.1cm]
$0.\phantom{00}<(NUV - r)_{\rm{abs,corr}} <1.35$& \phantom{0}4507 & 0.05 &1.79 $\pm$ 0.17& 1.55 $\pm$ 0.10& 0.45 $\pm$ 0.04\\[0.1cm]
$1.35<(NUV - r)_{\rm{abs,corr}} <1.64$ & \phantom{0}4503 & 0.05 & 2.07 $\pm$ 0.22 &  1.51 $\pm$ 0.10& 0.51 $\pm$ 0.04\\[0.1cm]
$1.64<(NUV - r)_{\rm{abs,corr}} <4.\phantom{00}$ & \phantom{0}4592 & 0.05 & 2.75 $\pm$ 0.16 &  1.54 $\pm$ 0.06& 0.62 $\pm$ 0.03\\[0.1cm]
\enddata
\end{deluxetable*}

%% file: tab4.tex
\begin{deluxetable*}{cccccccc}
\tablecolumns{6} \tabletypesize{\footnotesize} \tablewidth{0pt}
\tablecaption{\small MIS/SDSS sample $M_{\ast} - SSFR$
clustering results\tablenotemark{*}\label{tab_mis_m_ssfr}} \tablehead{
\multicolumn{2}{c}{Sample\tablenotemark{+}} & \colhead{$N_{gal}$}
&\colhead{$\langle z \rangle$} & \colhead{$\langle M_{\ast}\rangle [M_{\odot}]$}&\colhead{$r_0$ [$h^{-1}$ Mpc]} &
\colhead{$\gamma$} & \colhead{$b_8$}} \startdata
\multirow{3}{*}{$-10.\phantom{0}<\log(SFR/M_{\ast})<-9.$} & $10^{\phantom{0}8.\phantom{00}}<M_{\ast}<10^{\phantom{0}9.9\phantom{0}}$ &2103 & 0.04 & $10^{\phantom{0}9.56}$ &2.46 $\pm 0.21$ &1.6\phantom{0} $\pm$ 0.1\phantom{0} & 0.56 $\pm$ 0.04\\[0.1cm]
& $10^{\phantom{0}9.9\phantom{0}}<M_{\ast}<10^{10.48}$ &3347 & 0.07 &$10^{10.24}$ &3.58 $\pm 0.17$ &1.65 $\pm$ 0.07 & 0.76 $\pm$ 0.03\\[0.1cm]
& $10^{10.48}<M_{\ast}<10^{12.\phantom{00}}$ &3375 & 0.12 &$ 10^{10.82}$ &3.48 $\pm 0.16$ &1.65 $\pm$ 0.06 & 0.76 $\pm$ 0.03\\[0.1cm]
\multirow{2}{*}{$-11.<\log(SFR/M_{\ast})<-10.$} & $10^{\phantom{0}9.9\phantom{0}}<M_{\ast}<10^{10.48}$ &1142 & 0.06 &$10^{10.32}$ &5.23 $\pm 0.4\phantom{0}$ &1.60 $\pm$ 0.06 & 1.03 $\pm$ 0.06\\[0.1cm]
& $10^{10.48}<M_{\ast}<10^{12.\phantom{00}}$ &8461 & 0.11 &$10^{11.07}$ &5.10 $\pm 0.17$ &1.73 $\pm$ 0.03 & 1.06 $\pm$ 0.03\\[0.1cm]
\multirow{2}{*}{$-12.6<\log(SFR/M_{\ast})<-11.$} & $10^{\phantom{0}9.9\phantom{0}}<M_{\ast}<10^{10.48}$ &\phantom{0}304 & 0.05 &$10^{10.30}$ &4.95 $\pm 0.9\phantom{0}$ &1.95$^{+0.15}_{-0.05}$ & 1.04 $\pm$ 0.18\\[0.1cm]
& $10^{10.48}<M_{\ast}<10^{12.\phantom{00}}$ &2925 & 0.10 &$10^{11.23}$ &6.47 $\pm 0.29$ &1.93 $\pm$ 0.04 & 1.38 $\pm$ 0.06\\[0.1cm]
\enddata
\tablenotetext{*}{With dust extinction correction}
\tablenotetext{+}{The specific star formation rate, $SFR/M_{\ast}$ is
in units of $yr^{-1}$.}
\end{deluxetable*}